\documentclass[12pt]{article}
\usepackage[margin=2.5cm]{geometry}
\usepackage{bm}
\usepackage{comment}
\usepackage{xurl}
\usepackage{textcomp}
\usepackage{gensymb}

\usepackage{multirow}
\usepackage{graphicx}
\usepackage{amsmath}
\usepackage{amsfonts}
\usepackage{amssymb}
\usepackage{mathrsfs}
\usepackage{nicefrac}
\usepackage[authoryear]{natbib}

\newtheorem{theorem}{Theorem}
\newtheorem{corollary}{Corollary}

\title{Earth and space observations meet complex algebras: from complex to octonions for multivariate autoregressive time series analysis}

\author{
Susana Eyheramendy$^{a,b,c,*,\dagger}$,
Felipe Elorrieta$^{c,d,*}$,\\
Wilfredo Palma$^{c,*}$,
Javier Lopatin$^{a,b}$
\\[1.5ex]
\small $^{a}$Faculty of Engineering and Science, Universidad Adolfo Ib\'a\~nez,\\
\small Av. Diagonal Las Torres 2700, Pe\~nalol\'en, Santiago, RM, Chile\\
\small $^{b}$Data Observatory Foundation, ANID Technology Center No. DO210001, Santiago, RM, Chile\\
\small $^{c}$Millennium Institute of Astrophysics, ICM-ANID ICN12\_009, Santiago, Chile\\
\small $^{d}$Department of Mathematics and Computer Science, Faculty of Science,\\
\small Universidad de Santiago, Av. Libertador Bernardo O'Higgins 3663,\\
\small Estaci\'on Central, Santiago, RM, Chile\\[1ex]
\small $^{*}$S.E., F.E., and W.P. contributed equally to this work.\\
\small $^{\dagger}$Corresponding author: susana.eyheramendy@uai.cl
}

\date{}

\begin{document}

\maketitle

\begin{abstract}
Many scientific datasets arise as multivariate time series observed at irregular time intervals, particularly in Earth observation and astronomical measurements. Classical time-series models typically assume that time is discrete and observations occur at equally spaced intervals, limiting their ability to capture dynamics when observation gaps vary. Existing approaches to irregular sampling often rely on continuous-time formulations, which implicitly assume that observation intervals are sufficiently small. We address this limitation by introducing a discrete-time framework that directly accommodates irregular observation gaps while preserving multivariate dependence
structures.

In this work, we introduce a novel framework for the analysis of irregularly observed multivariate time series based on hypercomplex autoregressive processes. The proposed framework embeds multivariate observations into a hypercomplex algebraic structure, allowing multiple variables and their interactions to be represented within a single mathematical entity while explicitly incorporating irregular observation gaps. The resulting process admits a structured matrix representation, which enables the model to be expressed as a state-space system.

This formulation provides a convenient estimation methodology in which model parameters can be inferred using Kalman filtering techniques. The state-space representation allows efficient recursive estimation and prediction even in the presence of irregular sampling intervals.

The performance of the proposed methodology is illustrated through simulated data and applications to remote sensing and astronomical datasets, where observations naturally occur at nonuniform time intervals. The results demonstrate that the proposed approach effectively captures temporal dynamics and cross-variable interactions that are difficult to model using traditional time-series techniques.

The proposed framework provides a flexible and computationally tractable methodology for analyzing irregularly sampled multivariate time series and opens new possibilities for modeling complex observational data across a wide range of scientific disciplines.
\end{abstract}

\maketitle

\section{Introduction}\label{sec1}

Over the last few decades, the exponential growth in the availability of large-scale data has transformed nearly every area of scientific research. In particular, fields like astronomy and remote sensing have experienced a data revolution, driven by continuous, high-frequency observations that generate massive volumes of complex information. These developments demand robust analytical frameworks, among which the analysis of multivariate time series has emerged as a crucial methodology.

In astronomy, time-domain surveys such as the Zwicky Transient Facility (ZTF)\cite{Patterson_2018,Bellm_2018,Juric_2019} and the Vera C. Rubin Observatory’s Legacy Survey of Space and Time (LSST)\cite{Ivezic_2019} are fundamentally reshaping how we study the cosmos. These surveys produce streaming data on celestial objects, capturing variability over time in multiple photometric bands. The resulting multivariate time series contain rich information about transient phenomena such as supernovae, variable stars, and potential exoplanets. Analyzing these datasets requires sophisticated statistical and machine learning tools capable of handling irregular sampling, noise, and high dimensionality, while also allowing for classification, clustering, anomaly detection, and prediction.

Similarly, in the rapidly advancing field of remote sensing, satellite missions now deliver continuous Earth observation data across various spectral bands and spatial scales. This includes multi- and hyperspectral imagery capturing environmental variables such as vegetation indices, land surface temperature, atmospheric composition, and soil moisture—over time. The combination of spectral, spatial, and temporal dimensions creates high-dimensional multivariate time series datasets. Applications range from monitoring climate dynamics and natural disasters to tracking deforestation, urban growth, and agricultural cycles\cite{Fuentes_2024}.

What unites these domains is the need to detect and understand temporal dynamics across multiple interrelated variables. Whether observing a galaxy or a forest, the objective is to extract meaningful patterns from complex and often noisy time-evolving data. This requires not only advances in computational power and storage but also in the development of scalable and interpretable models. 

Moreover, both fields face similar challenges in terms of data quality, heterogeneity, missing observations, and the integration of domain knowledge. 

The convergence of astronomy and remote sensing through the methodological framework of multivariate time series analysis underscores a broader shift toward data-driven, cross-disciplinary science. As data continues to grow in volume and complexity, developing robust analytical tools for understanding dynamic systems will be critical for unlocking insights not only about our universe but also about our own planet.



In this article we present a general framework for multivariate time series models that we  implement in data from satellite  and astronomical surveys \cite{tsay_2010}. This framework defines autoregressive models for multivariate time series observed at irregular time intervals \cite{jones1981fitting,brockwell2001}. Specifically, these models are defined for $2$, $4$ and $8$ dimensional vector time series, but they can handle dimensions other than powers of $2$ by assuming that some of the multivariate time series coordinates are latent. Example of this are shown in Section \ref{sec:astro}. 

A key challenge faced by an univariate autoregressive model for a time series observed at irregular intervals is that the autocorrelation parameter needs to be evaluated at the power of, in principle, any positive rational  number. Specifically, if the autoregressive model is represented as

\begin{equation}\label{eq0}
 y_{t_s}=\phi^{t_s-t_{s-1}}y_{t_{s-1}}+\epsilon_{t_s}\mbox{ where }\epsilon_{t_s}\sim (0,\sigma^2_{t_s})
\end{equation}

\noindent
where $t_1,\ldots,t_n$ are the observational times and $\phi$ the autocorrelation parameter, then for any real value $\phi\in (-1,1)$, $\phi^{t_s-t_{s-1}}$ needs to be defined for any $\delta_s=t_s-t_{s-1}$, a positive rational number. But when $\phi$ is negative, the expression $\phi^{t_s-t_{s-1}}$ is defined only when $\delta_s=h$  is constant, otherwise it can not be defined.  Some models (e.g. the iAR \cite{Eyheramendy_2018} and CAR \cite{Kelly_2014}) restrict $\phi\in (0,1)$ to be always positive \cite{Kelly_2014,Eyheramendy_2018}, allowing $\delta_s$ to take any positive rational value. The standard autoregressive model (AR) restrict $\delta_s=h$ to be constant, allowing $\phi\in (-1,1)$ to be in the full range\cite{Brockwell_2016}.

The Complex irregular Autoregressive model (CiAR)  \cite{Elorrieta_2019} (introduced by our team) which uses complex numbers to define the model, enables $\phi\in (-1,1)$ for $\delta_s$ any positive rational number. The model is represented by the same equation 

\[ \bm{y}_{t_s}=\bm{\phi}^{t_s-t_{s-1}}\bm{y}_{t_{s-1}}+\bm{\epsilon}_{t_s}\]

\noindent
but in this model $\bm{y}_{t_s}=y_{t_s}^R+iy_{t_s}^I$ and $\bm{\epsilon}_{t_s}=\epsilon_{t_s}^R+i\epsilon_{t_s}^I$ are complex random variables and $\bm{\phi}=\phi^R+i\phi^I$ is a complex parameter. Through the polar representation of complex numbers and the de Moive's theorem \cite{Churchill_1990}, it is possible to write the model as follows,

\begin{equation}\label{eq1}
 \bm{y}_{t_s}=\bm{\alpha}_{t_s}\bm{y}_{t_{s-1}}+\bm{\epsilon}_{t_s}
\end{equation}

\noindent
which avoids having to exponentiate negative numbers. This is the key feature with which an autoregressive model can be defined for $\phi\in (-1,1)$  when observations occurred at irregular gaps. 
 We show in the next section how this model has been extended by our team to multivariate time series in $2$ dimensions, the BiAR \cite{Elorrieta_2021}. Further we show how the same framework allows for extensions to $2$, $4$ and $8$ dimensions.  
 
 Note that $\bm{\alpha}_{t_s}$ is also a complex number $\bm{\alpha}_{t_s}=\alpha_{t_s}^R+i\alpha_{t_s}^I$ where 
$\alpha_{t_s}^R=\|\phi\|^{\delta_{s}}\mbox{cos}(\delta_s\psi)$
and $\alpha_{t_s}^I=\|\phi\|^{\delta_{s}}\mbox{sin}(\delta_s\psi)$,
with $\psi=\mbox{arccos}(\frac{\phi^R}{\|\phi\|})$, $\|\phi\|=\sqrt{\phi^{R2}+\phi^{I2}}$ and $\delta_s=t_s-t_{s-1}$.

In Equation (\ref{eq1}) the complex parameter $\bm{\alpha}_{\delta_s}=\alpha_{\delta_s}^R+i\alpha_{\delta_s}^I$ is multiplied by the complex random variable $\bm{y}_{t_{s-1}}=y_{t_{s-1}}^R+iy_{t_{s-1}}^I$ from which a new complex is obtained:

\[\bm{\alpha}_{t_s}\bm{y}_{t_{s-1}}=(\alpha_{t_s}^{R}y_{t_{s-1}}^{R}-\alpha_{t_s}^{I}y_{t_{s-1}}^{I})+i(\alpha_{t_s}^{I}y_{t_{s-1}}^{R}+\alpha_{t_s}^{R}y_{t_{s-1}}^{I}).\]

\noindent
Equation (\ref{eq1}) can also be represented in two dimensions, with one dimension representing the real part and the other dimension representing the imaginary part of the complex sequence, in matrix notation it is as follows:

\begin{equation}
\label{eq2}
\left(\begin{array}{c} y_{t_s}^{R} \\  y_{t_s}^{I} \end{array} \right)= \left(\begin{array}{cc} \alpha_{t_s}^{R}  & -\alpha_{t_s}^{I} \\ \alpha_{t_s}^{I} & \alpha_{t_s}^{R} \end{array} \right)\left(\begin{array}{c} y_{t_{s-1}}^{R} \\  y_{t_{s-1}}^{I} \end{array} \right) + \left(\begin{array}{c} \varepsilon_{t_s}^R \\ \varepsilon_{t_s}^I \end{array} \right)
\end{equation}

Matrix Equation (\ref{eq2}) implicitly defines the transition equation in a state-space model representation. The complete representation requires an observation equation which can take different forms. For a univariate time series model, the following equation set the imaginary part as latent (i.e. this is the so-called CiAR model):

\begin{equation}  \label{eq3}
y_{t_s}^{R\ast}  = \left(\begin{array}{cc} 1 & 0 \end{array} \right) \left(\begin{array}{c} y_{s_j}^{R}  \\  y_{t_s}^{I}  \end{array} \right)
\end{equation}

\noindent
 But for a two dimensional time series model, the observation equation is the identity matrix: 

\begin{equation}  \label{eq4}
\left(\begin{array}{c} y_{t_s}^{R\ast} \\y_{t_s}^{I\ast} \end{array} \right) = \left(\begin{array}{cc} 1 & 0\\ 0 & 1 \end{array} \right) \left(\begin{array}{c} y_{s_j}^{R}  \\  y_{t_s}^{I}  \end{array} \right)
\end{equation}

These set of equations can be extended to allow measurement errors in the observations:

\begin{equation}  \label{eq5}
\left(\begin{array}{c} y_{t_s}^{R\ast} \\y_{t_s}^{I\ast} \end{array} \right) = \left(\begin{array}{cc} 1 & 0\\ 0 & 1 \end{array} \right) \left(\begin{array}{c} y_{s_j}^{R}  \\  y_{t_s}^{I}  \end{array} \right) + \left(\begin{array}{c} v_{t_s}^{R} \\v_{t_s}^{I} \end{array} \right).
\end{equation}

The random errors $v_{t_s}^{R}\mbox{ and } v_{t_s}^{I}$ have known variances and expected values equal to zero.
With the state-space representation of the model, i.e. (\ref{eq2}) + (\ref{eq4}) or  (\ref{eq2}) + (\ref{eq5}), it is possible to implement a Kalman filter algorithm to estimate the model parameters which enables to do forecasting and prediction \cite{kalman1960,durbin2012,harvey1990,shumway2017}.

Note that the model specified by Equation (\ref{eq2}) assumes that the autocorrelations of both univariate sequences are the same and the cross-correlation are the same in magnitude but have different sign at each observational time. Therefore, instead of having four free parameters for the autocorrelations and cross-correlations, the model has two parameters. This poses some restrictions in the specification of the model, but at the same time is less expensive on data as it requires a smaller number of parameters to be estimated.

Summarizing, we have presented a model with complex random variables that can be represented as a two dimensional model for time series that can be written as a state-space system which in turn enables estimation throw a Kalman filter algorithm. This model fits time series sequences observed at irregular time intervals and allows for the estimation of positive as well as negative autocorrelations, which in general is not possible (e.g. \cite{Eyheramendy_2018,Kelly_2014}).  Nevertheless, this model has some strong assumptions regarding the autocorrelations of the univariate series and the cross-correlations between the sequences. The CiAR \cite{Elorrieta_2019} (represented by Equations (3) + (4)) and the BiAR \cite{Elorrieta_2021}  (represented by Equations (3) + (5) or (3)+(6)) models have been published already, as well as a univariate version of the model, i.e iAR model \cite{Eyheramendy_2018}. 

In this article we aim to extend these models to allow for higher dimensional time series analysis, and to describe a general framework from which these models have been generated. There is a direct extension of the BiAR model to a $2$, $4$ or $8$ dimensional vector-valued time series model by replacing complex random variables in Equation (\ref{eq0}) with generalized complex, quaternion or octonion random variables \cite{baez2002,conway2003}. We call the model for four dimensional time series the QiAR, as it is based on quaternion random variables, and we call OiAR the model for eight dimensional  time series as it is based on octonion random variables. In what follows we formulate the general model and its properties with proofs shown in the Appendix. The QiAR model is presented in detail in Section \ref{sec:qiar}, a simulation study and real data applications are shown in Section \ref{sec:results}. Some properties of the complex, quaternion and octonion algebras as well as details of the specification of the OiAR model are presented in the Appendix.

Although the general theoretical framework developed here accommodates
multivariate time series of dimension up to eight through the octonion
algebra (the OiAR model), the empirical results in this article focus
exclusively on the four-dimensional QiAR model. The univariate, bivariate
and complex versions of this framework -- the iAR, BiAR and CiAR models,
respectively -- have already been published [10, 13, 15]. Including a full
simulation study and set of applications for the octonion-based OiAR model
in this article would considerably lengthen the manuscript. We therefore
present the general OiAR formulation and its theoretical properties in the
Appendix, and reserve a dedicated empirical treatment of the OiAR model,
together with its own simulation study and applications, for a separate
article, so that this eight-dimensional extension receives the attention it
warrants.

\subsection*{The Model}\label{sec:themodel}
An auto-regressive dependency is the basic structure of the models that we are proposing. For given observational times $t_1,\ldots,t_n$, such that $t_j-t_{j-1}=\delta_j\in \mathbb{Q}$ are rational numbers and generally not equal, the models can be written as follows:\\

\begin{equation}
\bm{y}_{t_j}=\bm{\phi}^{t_j-t_{j-1}}\bm{y}_{t_{j-1}} +\bm{\epsilon_{t_{j}}}\label{eq:its}
\end{equation}
\vspace{0.2in}

\noindent
where $\mbox{E}(\bm{\epsilon_{t_{j}}})=\bm{0}$ and $\mbox{Var}(\bm{\epsilon_{t_{j}}})=\bm{\sigma^2_{t_j}}$. The random variables (i.e. $\bm{y}_{t_j},  \bm{\epsilon}_{t_{j}}$) and parameters (i.e. $\bm{\phi}, \bm{\sigma}_{t_{j}}$) in eq. (\ref{eq:its}) can all take values in either the complex ($\mathbb{C}$), the quaternions ($\mathbb{H}$) or the octonions ($\mathbb{O}$), depending on the dimension of the time series that are being modeled. In other words, for a two dimensional time series $\bm{y}_{t_j}, \bm{\phi}, \bm{\sigma}_{t_{j}}, \bm{\epsilon}_{t_{j}}\in \mathbb{C}$, for a four dimensional time series $\bm{y}_{t_j}, \bm{\phi}, \bm{\sigma}_{t_{j}}, \bm{\epsilon}_{t_{j}}\in \mathbb{H}$ and for an eight dimensional time series $\bm{y}_{t_j}, \bm{\phi}, \bm{\sigma}_{t_{j}}, \bm{\epsilon}_{t_{j}}\in \mathbb{O}$. This framework also allows to adjust time series whose dimension are not power of $2$, e.g. a three dimensional time series can be modeled by a quaternion representation that assumes a latent time series in one of the coordinates. Another example is the CIAR model. This model corresponds to a BIAR model \cite{Elorrieta_2021} that assumes that the imaginary sequence of the complex series is latent. Further, a two dimensional time series can be modeled using  complex random variables as well as, for instance, quaternions. This latter representation will allow some extra flexibility as it will be shown in what follows. In general, a model represented with complex numbers can fit univariate time series as well as bivariate time series; a model represented with quaternion random variables can fit multivariate time series of dimension $2$, $3$ or $4$; and a model represented with octonion random variables can fit multivariate time series of dimension $2,\ldots, 8$.

This upper bound of eight dimensions is not a modeling choice but a
mathematical necessity. By Hurwitz's theorem, the real numbers, the
complex numbers, the quaternions and the octonions are the only normed
division algebras \cite{baez2002}; no analogous algebra exists in any other
dimension. Since Theorem 1 relies on a polar (Euler/de Moivre)
representation that requires a multiplicative norm, the framework
presented here cannot be extended to construct an analogous iAR-type
model beyond dimension 8 without giving up the algebraic properties
(norm, alternativity) that make the exponentiation in Equation (7)
well defined.

Some properties of these models are stated in the following theorems, with proofs in the Appendix. \\

\begin{theorem}[Moivre] The process described by Equation (\ref{eq:its}) can be written as\\

\begin{equation}
\bm{y}_{t_j}=\bm{\alpha}_{t_j}\bm{y}_{t_{j-1}} +\bm{\epsilon_{t_{j}}}\label{eq:itsMoivre}
\end{equation}

\noindent
when the random variables and parameters of Equation (\ref{eq:its})  take values in $\mathbb{C}$, $\mathbb{H}$ or $\mathbb{O}$ (or their generalized versions $\mathbb{C}_{\alpha}$, $\mathbb{H}_{\alpha\beta}$ or $\mathbb{O}_{\alpha\beta\gamma}$).
\end{theorem}

\begin{theorem}[State-space representation] The process described by Equation (\ref{eq:itsMoivre}) can be represented as a state-space system as follows:\\

\begin{eqnarray}\label{eq:itsSS}
\bm{x}_{t_j} &=&F_{t_j}\bm{x}_{t_{j-1}}+V_{t_j}\\
\bm{y}_{t_j} &=&G_{t_j}\bm{x}_{t_{j}}+W_{t_j} \mbox{ j=1,\ldots,n}
\end{eqnarray}

\noindent
where $\bm{y}_{t_j}$ is a random vector of dimensions $d$ with $d\leq 8$. The matrices $F_{t_j}$ are of dimension $2\times 2$, $4\times 4$ or $8\times 8$  depending on whether the model is defined in the complex, quaternions or octonions, respectively. and the matrices $G_{t_j}$ are of dimension $p\times 2$, $q\times 4$ or $r\times 8$ with $p\leq 2$, $q\leq 4$ and $r\leq 8$, depending on how many observed sequences are available.
\end{theorem}

\begin{corollary}
Multivariate time series of order $p$, with $p<8$ not having a power of two, can also be analyzed using the model described in Eq. (\ref{eq:its}). In particular, The BIAR model can be extended  by considering different sets of two-coordinates from the four coordinates of the QiAR model.
\end{corollary}



\begin{theorem}[Existence of a solution]
The process defined by Equation (\ref{eq:its}), has a solution given by 
\[\bm{y}_{t_j}=\sum_{k=0}^{\infty}\bm{\phi}^{t_j-t_{j-k}}\bm{\epsilon}_{t_{j-k}}\]

\noindent
and the variance of the process is finite with  $Var(\bm{y}_{t_j})=\Sigma$.

\end{theorem}

\begin{theorem}[Stability] The process defined by equation \eqref{eq:its} is stable if

\begin{equation}
    \| \mathbf{\phi} \|_L <1
\end{equation}

\noindent
where the norm $ \| . \|_L$ corresponds to the $L_2$ (i.e. the Euclidean norm) in the BiAR, QiAR and OiAR models but for the generalized version of the BiAR, QiAR and OiAR, this norm is the corresponding for generalized complex, quaternions and octonions (detailed in the Appendix).

\end{theorem}

In what follows we detailed the QiAR model. The OiAR model is similarly generated, and the details are described in the Appendix.

\section{The QiAR model}\label{sec:qiar}

Based on Theorem $1$ and $2$ the general model \\

\[\bm{y}_{t_s}=\bm{\phi}^{t_s-t_{s-1}}\bm{y}_{t_{s-1}} +\bm{\epsilon_{t_{s}}},\]
\vspace{0.2in}

\noindent
for quaternion random variables $\bm{y}_{t_{j}}$ and $\bm{\epsilon_{t_{j}}}$ and quaternion parameter $\bm{\phi}$, can be expressed in matrix notation as a state-space system as follows:

\begin{eqnarray*}
\bm{x}_{t_s} &=&F_{t_s}\bm{x}_{t_{s-1}}+V_{s_j}\\
\bm{y}_{t_s} &=&G_{t_s}\bm{x}_{t_{s}}+W_{s_j} \mbox{ j=1,\ldots,n}
\end{eqnarray*}

\noindent
where $F_{t_s}$ is defined as

\begin{equation}\label{eq:Fquatgen1}
F_{t_s} = \left(\begin{array}{cccc} \alpha_{t_s}^{(0)} & -\alpha\alpha_{t_s}^{(1)}  & -\beta\alpha_{t_s}^{(2)} & -\alpha\beta\alpha_{t_s}^{(3)} \\ 
\alpha_{t_s}^{(1)} & \alpha_{t_s}^{(0)} & -\beta\alpha_{t_s}^{(3)} & \beta\alpha_{t_s}^{(2)} \\ 
\alpha_{t_s}^{(2)} & \alpha\alpha_{t_s}^{(3)} & \alpha_{t_s}^{(0)} & -\alpha\alpha_{t_s}^{(1)}  \\ 
\alpha_{t_s}^{(3)} & -\alpha_{t_s}^{(2)} & \alpha_{t_s}^{(1)}  & \alpha_{t_s}^{(0)} \end{array} \right).\end{equation} 
\vspace{0.2in}

For $\alpha=1$ and $\beta=1$, these previous equations correspond to the model that is based on the so-called Hamiltonian quaternions. But $\alpha$ and $\beta$ can take other values, and when this is the case, the model is based on the generalized quaternions. In this article we allow $\alpha$ and $\beta$ to take values in $\{-1,0,1\}$. Therefore, the QiAR has four parameters, $\phi^{(0)}, \phi^{(1)}, \phi^{(2)}, \phi^{(3)}$ that defines the coordinates of the quaternion $\bm{\phi}$, plus $\alpha$ and $\beta$ to estimate the autocorrelation and cross-correlation in the multivariate time series.

The observational matrix $G_{t_s}$ will be the four-dimensional identity matrix for four-dimensional multivariate time series, but otherwise can take other dimensions. For instance, in the example given in Section 3.3, this matrix was defined for a two-dimensional multivariate time series.

\section{Results}\label{sec:results}

\subsection{Simulations}
\label{sec:simhq}
Several simulations were performed with the following aims: i) to assess the precision and potential bias of the parameter's estimators; ii) to assess the appropriateness of the algorithm at estimating and recovering the parameters with which the time series were generated;  iii) to compare the model  with the gold standard VAR model \cite{sims1980macroeconomics,hamilton1994time,lutkepohl2005new}; iv) to quantify the quality of the model for interpolation; v) to quantify the quality of the model at forecasting.

\subsubsection{Parameter estimation in time series with regular and irregular observational time-gaps}
\label{sec:sim1}
The first simulation experiments aim to verify that the proposed maximum likelihood estimation method for the QiAR model is accurate. For this purpose, $200$ simulations of the QiAR process were generated independently for each value of the parameter vector $\phi=(\phi^{(0)},\phi^{(1)},\phi^{(2)},\phi^{(3)})^T$ that defines a process of length $n=300$. Note that the parameter vector  was chosen so that the determinant of the transition matrix would take values in the whole range between 0 and 1. For each,  we obtained the maximum likelihood estimator. Table \ref{t1} shows the mean estimate and its standard error when the QiAR processes were generated using regular times.  Table \ref{t2} shows these outcomes when the QiAR processes were simulated using irregular times. The time gaps were randomly generated from a uniform distribution in the range $(1,5)$. Note in the tables that all four parameters are accurately estimated by the proposed maximum likelihood estimation procedure for both regular and irregular times. 

\begin{table}[ht]
\centering
\begin{tabular}{r|rrrr|rrrr}
\hline
detF & $\phi^{(0)}$ & $\phi^{(1)}$ & $\phi^{(2)}$ & $\phi^{(3)}$ &  $\hat{\phi}^{(0)}$ & $\hat{\phi}^{(1)}$ & $\hat{\phi}^{(2)}$ & $\hat{\phi}^{(3)}$\\
\hline
0.040 & 0.300 & 0.100 & 0.300 & 0.100 & 0.301 & 0.099 & 0.307 & 0.099 \\
      &       &       &       &       & (0.020) & (0.031) & (0.018) & (0.020) \\
0.053 & -0.300 & -0.300 & -0.100 & 0.200 & -0.296 & -0.306 & -0.102 & 0.204 \\
      &        &        &        &       & (0.027) & (0.026) & (0.027) & (0.023) \\
0.185 & -0.300 & 0.300 & -0.300 & 0.400 & -0.302 & 0.300 & -0.307 & 0.401 \\
      &        &       &        &       & (0.020) & (0.022) & (0.020) & (0.022) \\
0.260 & 0.400 & -0.500 & -0.300 & -0.100 & 0.404 & -0.500 & -0.295 & -0.105 \\
      &       &        &        &        & (0.021) & (0.019) & (0.023) & (0.024) \\
0.360 & 0.200 & 0.400 & 0.600 & -0.200 & 0.201 & 0.396 & 0.605 & -0.198 \\
      &       &       &       &        & (0.018) & (0.016) & (0.012) & (0.016) \\
0.490 & 0.800 & 0.200 & -0.100 & 0.100 & 0.800 & 0.199 & -0.093 & 0.108 \\
      &       &       &        &       & (0.010) & (0.013) & (0.018) & (0.019) \\
0.608 & -0.800 & 0.100 & 0.300 & -0.200 & -0.802 & 0.101 & 0.301 & -0.194 \\
      &        &       &       &        & (0.006) & (0.015) & (0.015) & (0.014) \\
0.740 & -0.400 & 0.300 & 0.600 & -0.500 & -0.402 & 0.294 & 0.600 & -0.503 \\
      &        &       &       &        & (0.008) & (0.010) & (0.008) & (0.010) \\
0.828 & 0.800 & 0.100 & -0.500 & 0.100 & 0.800 & 0.097 & -0.500 & 0.099 \\
      &       &       &        &       & (0.006) & (0.011) & (0.006) & (0.008) \\
0.903 & 0.100 & -0.600 & 0.700 & 0.300 & 0.101 & -0.603 & 0.699 & 0.297 \\
      &       &        &       &       & (0.008) & (0.006) & (0.005) & (0.006) \\
0.960 & -0.300 & 0.800 & 0.300 & 0.400 & -0.301 & 0.799 & 0.301 & 0.401 \\
      &        &       &       &       & (0.004) & (0.003) & (0.005) & (0.004) \\
\hline
\end{tabular}
\caption{Maximum likelihood (ML) estimation of the parameters $\phi^{(0)}$, $\phi^{(1)}$, $\phi^{(2)}$, $\phi^{(3)}$ of the QiAR model generated with regular times and sample size $n=300$. The first column shows the determinant of the transition matrix for each parameter set (columns 2-5). The estimators of the parameters are in columns 6-9, and their standard deviations, estimated by the Monte Carlo experiments of each ML estimate, is shown in parentheses. }
\label{t1}
\end{table}

\begin{table}[ht]
\centering
\begin{tabular}{r|rrrr|rrrr}
\hline
detF & $\phi^{(0)}$ & $\phi^{(1)}$ & $\phi^{(2)}$ & $\phi^{(3)}$ & $\hat{\phi}^{(0)}$ & $\hat{\phi}^{(1)}$ & $\hat{\phi}^{(2)}$ & $\hat{\phi}^{(3)}$\\
\hline
0.040 & 0.300 & 0.100 & 0.300 & 0.100 & 0.293 & 0.120 & 0.292 & 0.115 \\
      &       &       &       &       & (0.032) & (0.049) & (0.042) & (0.067) \\
0.053 & -0.300 & -0.300 & -0.100 & 0.200 & -0.310 & -0.324 & -0.066 & 0.185 \\
      &        &        &        &       & (0.041) & (0.088) & (0.162) & (0.187) \\
0.185 & -0.300 & 0.300 & -0.300 & 0.400 & -0.301 & 0.287 & -0.298 & 0.394 \\
      &        &       &        &       & (0.027) & (0.062) & (0.051) & (0.049) \\
0.260 & 0.400 & -0.500 & -0.300 & -0.100 & 0.401 & -0.500 & -0.299 & -0.106 \\
      &       &        &        &        & (0.015) & (0.027) & (0.035) & (0.036) \\
0.360 & 0.200 & 0.400 & 0.600 & -0.200 & 0.200 & 0.391 & 0.604 & -0.201 \\
      &       &       &       &        & (0.010) & (0.035) & (0.021) & (0.041) \\
0.490 & 0.800 & 0.200 & -0.100 & 0.100 & 0.800 & 0.198 & -0.105 & 0.104 \\
      &       &       &        &       & (0.008) & (0.011) & (0.014) & (0.013) \\
0.608 & -0.800 & 0.100 & 0.300 & -0.200 & -0.801 & 0.096 & 0.300 & -0.199 \\
      &        &       &       &        & (0.007) & (0.009) & (0.012) & (0.012) \\
0.740 & -0.400 & 0.300 & 0.600 & -0.500 & -0.397 & 0.305 & 0.597 & -0.502 \\
      &        &       &       &        & (0.004) & (0.019) & (0.015) & (0.013) \\
0.828 & 0.800 & 0.100 & -0.500 & 0.100 & 0.800 & 0.099 & -0.501 & 0.099 \\
      &       &       &        &       & (0.003) & (0.007) & (0.004) & (0.012) \\
0.903 & 0.100 & -0.600 & 0.700 & 0.300 & 0.099 & -0.605 & 0.695 & 0.305 \\
      &       &        &       &       & (0.003) & (0.012) & (0.010) & (0.015) \\
0.960 & -0.300 & 0.800 & 0.300 & 0.400 & -0.300 & 0.800 & 0.303 & 0.398 \\
      &        &       &       &       & (0.003) & (0.006) & (0.009) & (0.008) \\
\hline
\end{tabular}
\caption{Maximum likelihood (ML) estimation of the parameters $\phi^{(0)}$, $\phi^{(1)}$, $\phi^{(2)}$, $\phi^{(3)}$ of the QiAR model generated with irregular times and sample size $n=300$. The first column shows the determinant of the transition matrix for each parameter set (columns 2-5). The estimators of the parameters are in columns 6-9, and their standard deviations, estimated by the Monte Carlo experiments of each ML estimate, is shown in parentheses.}
\label{t2}
\end{table}

\subsection{Estimation of $\alpha$ and $\beta$}

In Section \ref{sec:simhq} we show an accurate parameter estimation for the QiAR model defined under Hamiltonian quaternions (i.e. $\alpha=1$ and $\beta=1$). In this section, we are interested in assessing whether this estimation accuracy holds when we consider other values of $\alpha$ and $\beta$. Furthermore, we are also interested in assessing whether we can recover the values of $\alpha$ and $\beta$ used to perform each experiment. For this purpose, we performed nine Monte Carlo experiments with 100 repetitions each. For each experiment, we generated a QiAR model of length $n=300$ with parameters $\phi^{(0)}=0.5$, $\phi^{(1)}=0.3$, $\phi^{(2)}=0.3$ and $\phi^{(3)}=0.2$, while the $\alpha$ and $\beta$ parameters are chosen according to a grid of points taking values in $\{-1, 0, 1\}$. Table \ref{tab:alpha} shows that the precise parameter estimation holds for any value of $\alpha$ and $\beta$, but with a higher standard deviation for the case where $\alpha$ or $\beta$ are negative. Furthermore, the estimation algorithm detects the correct value of $\alpha$ and $\beta$ in most of the simulation experiments performed, with an efficiency close to 100\% when the original parameter is equal to one.

\begin{table}[ht]
\centering
\scriptsize
\begin{tabular}{cccccccc}
\hline
$\alpha$ & $\beta$ & $\hat{\phi}^{(0)}$ & $\hat{\phi}^{(1)}$ & $\hat{\phi}^{(2)}$ & $\hat{\phi}^{(3)}$ & $\hat{\alpha}$ & $\hat{\beta}$\\
\hline
\textbf{1.0} & \textbf{1.0} &
\begin{tabular}{c}0.499\\{\scriptsize(0.023)}\end{tabular} &
\begin{tabular}{c}0.301\\{\scriptsize(0.038)}\end{tabular} &
\begin{tabular}{c}0.298\\{\scriptsize(0.035)}\end{tabular} &
\begin{tabular}{c}0.207\\{\scriptsize(0.044)}\end{tabular} &
\begin{tabular}{c}0.970\\{\scriptsize(0.119)}\end{tabular} &
\begin{tabular}{c}0.980\\{\scriptsize(0.099)}\end{tabular} \\
\textbf{1.0} & \textbf{0.5} &
\begin{tabular}{c}0.503\\{\scriptsize(0.023)}\end{tabular} &
\begin{tabular}{c}0.295\\{\scriptsize(0.035)}\end{tabular} &
\begin{tabular}{c}0.291\\{\scriptsize(0.050)}\end{tabular} &
\begin{tabular}{c}0.202\\{\scriptsize(0.045)}\end{tabular} &
\begin{tabular}{c}0.980\\{\scriptsize(0.099)}\end{tabular} &
\begin{tabular}{c}0.545\\{\scriptsize(0.144)}\end{tabular} \\
\textbf{1.0} & \textbf{0.0} &
\begin{tabular}{c}0.499\\{\scriptsize(0.023)}\end{tabular} &
\begin{tabular}{c}0.302\\{\scriptsize(0.030)}\end{tabular} &
\begin{tabular}{c}0.306\\{\scriptsize(0.045)}\end{tabular} &
\begin{tabular}{c}0.207\\{\scriptsize(0.045)}\end{tabular} &
\begin{tabular}{c}0.980\\{\scriptsize(0.099)}\end{tabular} &
\begin{tabular}{c}0.005\\{\scriptsize(0.050)}\end{tabular} \\
\textbf{0.5} & \textbf{1.0} &
\begin{tabular}{c}0.497\\{\scriptsize(0.024)}\end{tabular} &
\begin{tabular}{c}0.307\\{\scriptsize(0.049)}\end{tabular} &
\begin{tabular}{c}0.303\\{\scriptsize(0.035)}\end{tabular} &
\begin{tabular}{c}0.194\\{\scriptsize(0.044)}\end{tabular} &
\begin{tabular}{c}0.515\\{\scriptsize(0.166)}\end{tabular} &
\begin{tabular}{c}0.970\\{\scriptsize(0.119)}\end{tabular} \\
\textbf{0.5} & \textbf{0.5} &
\begin{tabular}{c}0.504\\{\scriptsize(0.028)}\end{tabular} &
\begin{tabular}{c}0.300\\{\scriptsize(0.043)}\end{tabular} &
\begin{tabular}{c}0.293\\{\scriptsize(0.049)}\end{tabular} &
\begin{tabular}{c}0.189\\{\scriptsize(0.056)}\end{tabular} &
\begin{tabular}{c}0.530\\{\scriptsize(0.139)}\end{tabular} &
\begin{tabular}{c}0.510\\{\scriptsize(0.174)}\end{tabular} \\
\textbf{0.5} & \textbf{0.0} &
\begin{tabular}{c}0.504\\{\scriptsize(0.028)}\end{tabular} &
\begin{tabular}{c}0.296\\{\scriptsize(0.039)}\end{tabular} &
\begin{tabular}{c}0.291\\{\scriptsize(0.050)}\end{tabular} &
\begin{tabular}{c}0.205\\{\scriptsize(0.061)}\end{tabular} &
\begin{tabular}{c}0.545\\{\scriptsize(0.144)}\end{tabular} &
\begin{tabular}{c}0.020\\{\scriptsize(0.099)}\end{tabular} \\
\textbf{0.0} & \textbf{1.0} &
\begin{tabular}{c}0.500\\{\scriptsize(0.027)}\end{tabular} &
\begin{tabular}{c}0.301\\{\scriptsize(0.051)}\end{tabular} &
\begin{tabular}{c}0.300\\{\scriptsize(0.029)}\end{tabular} &
\begin{tabular}{c}0.204\\{\scriptsize(0.048)}\end{tabular} &
\begin{tabular}{c}0.015\\{\scriptsize(0.086)}\end{tabular} &
\begin{tabular}{c}0.975\\{\scriptsize(0.110)}\end{tabular} \\
\textbf{0.0} & \textbf{0.5} &
\begin{tabular}{c}0.501\\{\scriptsize(0.031)}\end{tabular} &
\begin{tabular}{c}0.306\\{\scriptsize(0.049)}\end{tabular} &
\begin{tabular}{c}0.294\\{\scriptsize(0.038)}\end{tabular} &
\begin{tabular}{c}0.193\\{\scriptsize(0.050)}\end{tabular} &
\begin{tabular}{c}0.025\\{\scriptsize(0.110)}\end{tabular} &
\begin{tabular}{c}0.535\\{\scriptsize(0.147)}\end{tabular} \\
\hline
\end{tabular}
\caption{Maximum likelihood estimation of the parameters $\phi^{(0)}$, $\phi^{(1)}$, $\phi^{(2)}$, $\phi^{(3)}$, $\alpha$ and $\beta$ of the QiAR model generated with irregular times, sample size $n=300$ and parameters set as $\phi^{(0)}=0.5$, $\phi^{(1)}=0.3$, $\phi^{(2)}=0.3$ and $\phi^{(3)}=0.2$. The standard deviation estimated by the Monte Carlo experiments of each ML estimate is shown in parentheses. \label{tab:alpha}}
\end{table}

\subsubsection{Comparison with VAR(1) model}

\begin{figure*}
\begin{center}
\begin{minipage}{0.8\linewidth}
\centering
\includegraphics[width=\textwidth]{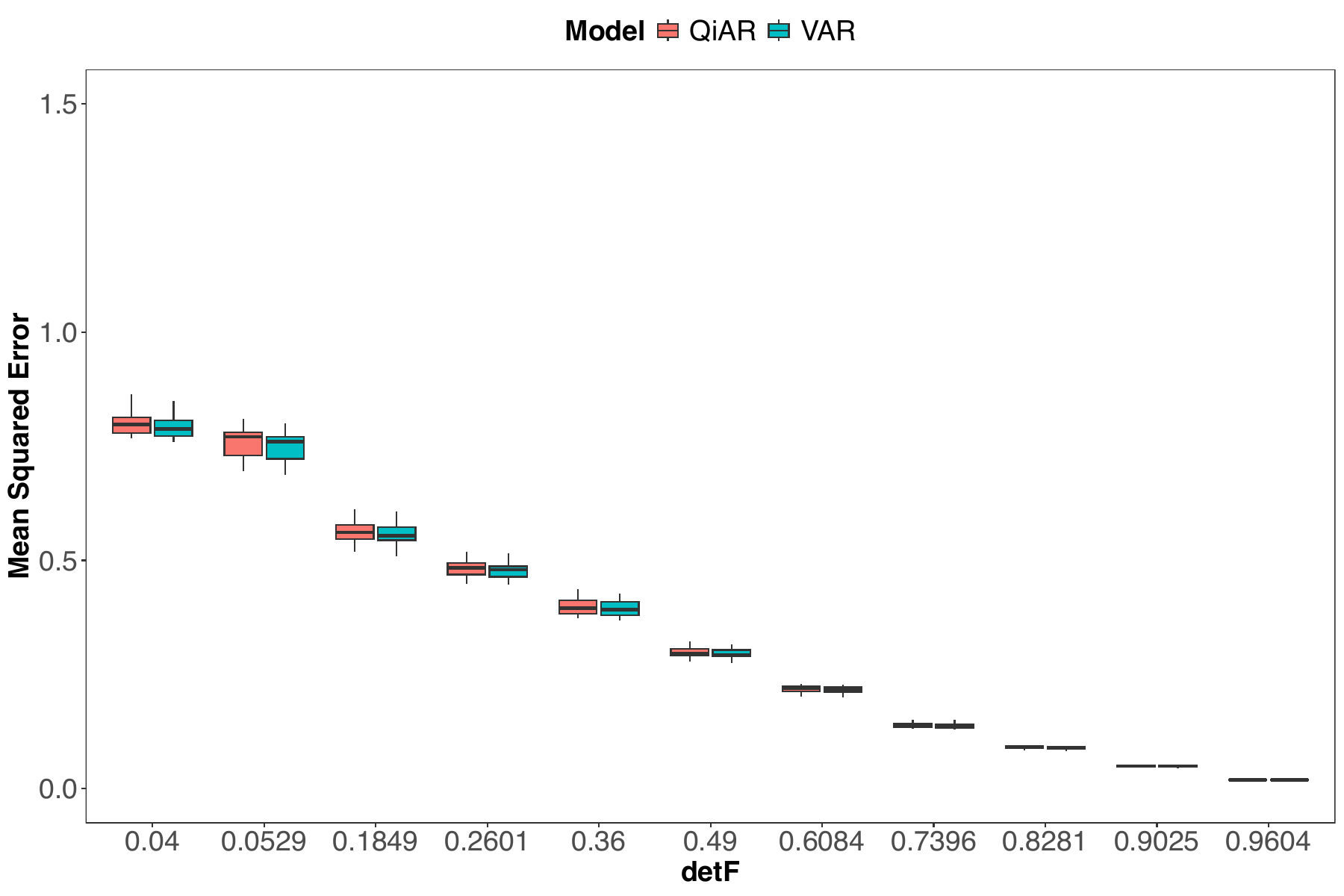}
\end{minipage}
\begin{minipage}{0.8\linewidth}
\centering
\includegraphics[width=\textwidth]{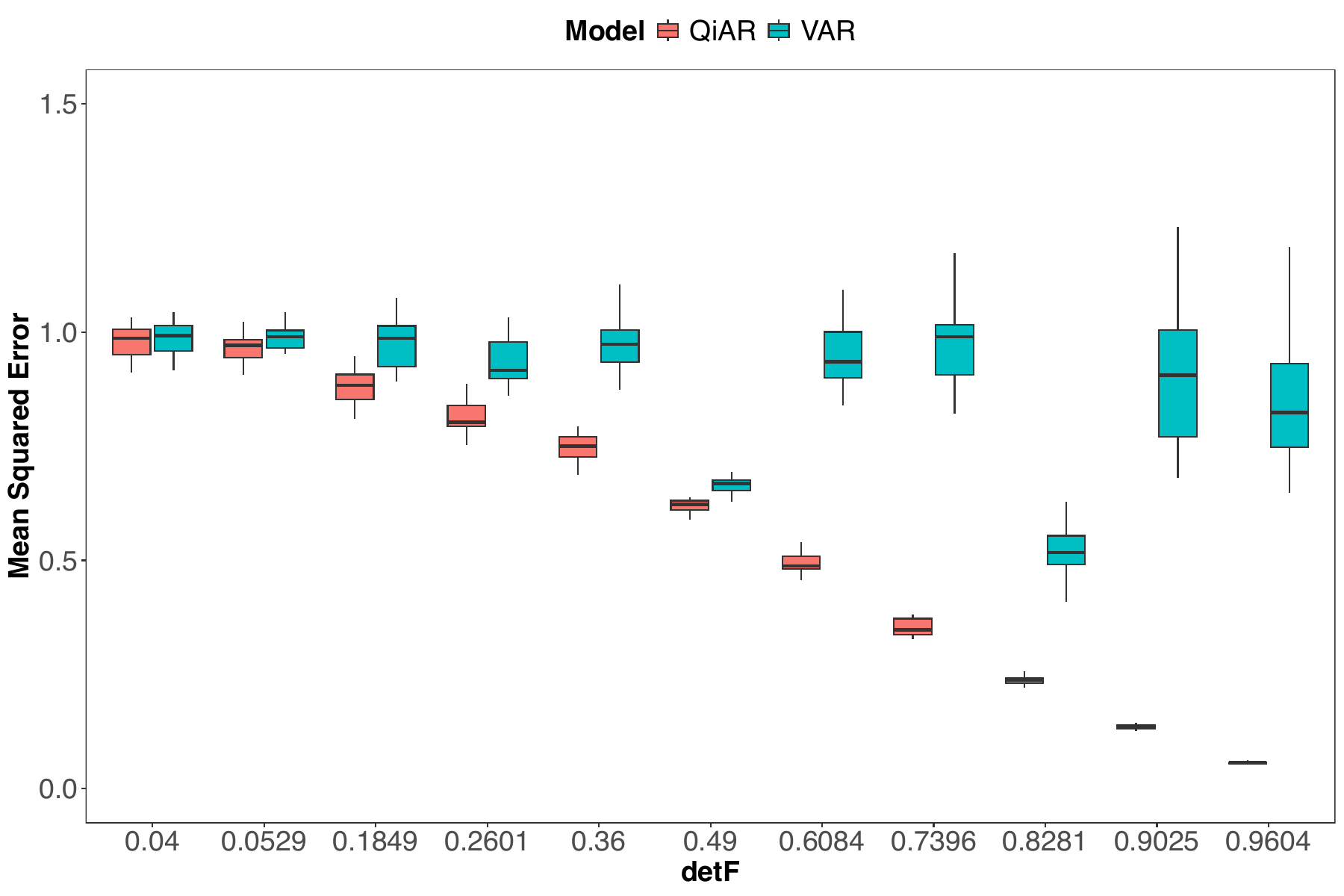}
\end{minipage}
\caption{Mean Squared Error Comparison for QiAR and VAR(1) models on simulated QiAR processes. Figure a) shows the results for regular times. Figure b) shows the results for irregular times. \label{fig:EPSQIAR2}}
\end{center}
\end{figure*}

The standard multivariate autoregressive model for regularly observed data is the vector autoregressive (VAR) model. The VAR model of order one (VAR(1)) for a multivariate time series of dimension $p=4$ is defined from the following equation:

\begin{equation}
y_t = \left(\begin{array}{c} y_{1t} \\ y_{2t} \\ y_{3t} \\ y_{4t}\end{array}\right) = \left(\begin{array}{cccc}
\phi_{11}& \phi_{12} & \phi_{13}& \phi_{14}\\ \phi_{21}& \phi_{22} & \phi_{23}& \phi_{24} \\ \phi_{31}& \phi_{32} & \phi_{33}& \phi_{34} \\ \phi_{41}& \phi_{42} & \phi_{43}& \phi_{44} \end{array}\right) \left(\begin{array}{c} y_{1,t-1} \\ y_{2,t-1} \\ y_{3,t-1} \\ y_{4,t-1} \end{array}\right)+ \left(\begin{array}{c}
\epsilon_{1t} \\ \epsilon_{2t} \\ \epsilon_{3t} \\ \epsilon_{4t}\end{array}\right)
\end{equation}

\noindent
where the parameters $\phi_{jj}, j=1,\ldots,4$ are the autocorrelation coefficients for each time series $y_{jt}$, and the parameters $\phi_{ij}, i,j=1,\ldots,4$ and $i \neq j$ are the cross-correlations coefficients. Note that the VAR model needs $16$ parameters to model the joint dependence of $4$ time series while the QiAR model requires estimating $6$ parameters. Furthermore, the VAR model is defined for regularly observed series. To quantify the impact of using the VAR model on irregularly observed data we performed the following experiment. 
For the simulated QiAR processes (see Section \ref{sec:sim1}), we fit both the QiAR and VAR(1) model and compute the mean squared error (MSE). Figure \ref{fig:EPSQIAR2} shows the MSE computed with the QiAR and VAR(1) model for each set of parameters both for regular and irregular times. Note that for regular times, both models make the same error. However, for irregular times, the QiAR model decreases its error as the determinant of the transition matrix increases. While the VAR(1) model maintains a stable MSE around one.\\

\subsection{Interpolation with the QiAR model}

We assessed the QiAR model's interpolation performance in a Monte Carlo
study with $m=100$ repetitions and sample size $n=300$, using irregular
time gaps drawn from a discrete uniform distribution on $\{1,\dots,5\}$.
Two parameter configurations were considered:
$\phi=(0.8,0.4,0.2,0.2)$ and $\phi=(0.5,0.3,0.3,0.2)$. For each, we
removed and then imputed three values per series, imputing either one
series at a time or all four simultaneously. Removed observations were
chosen so that the  adjacent time gap --- $\Delta_j = t_j -
t_{j-1}$ and $\Delta_{j+1} = t_{j+1} - t_j$ --- equaled 1, 3, or 5,
letting us test how temporal proximity affects accuracy.

\begin{center}
\begin{figure*}
\begin{minipage}{0.49\linewidth}
\centering
\includegraphics[width=\textwidth]{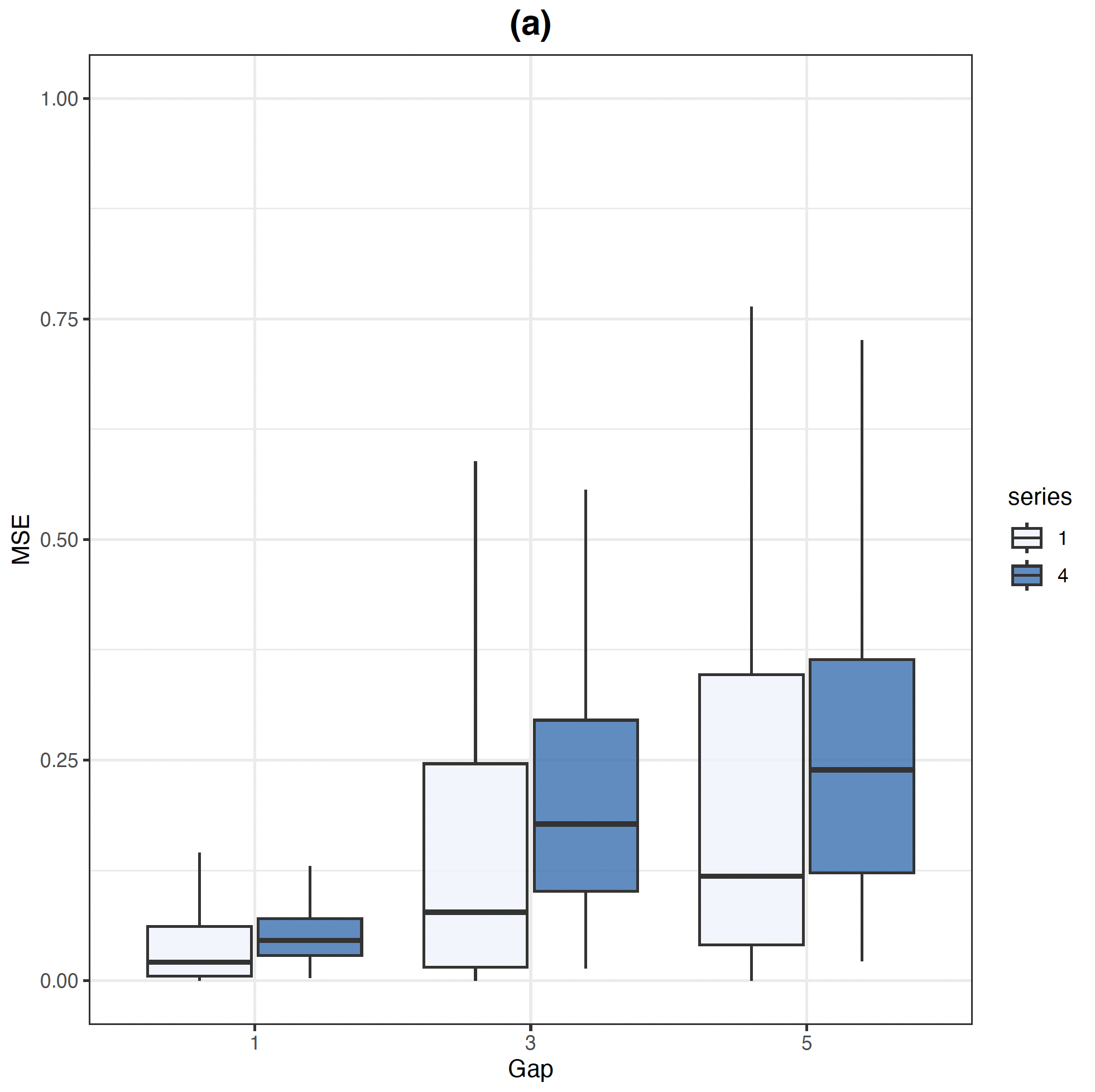}
\end{minipage}
\begin{minipage}{0.49\linewidth}
\centering
\includegraphics[width=\textwidth]{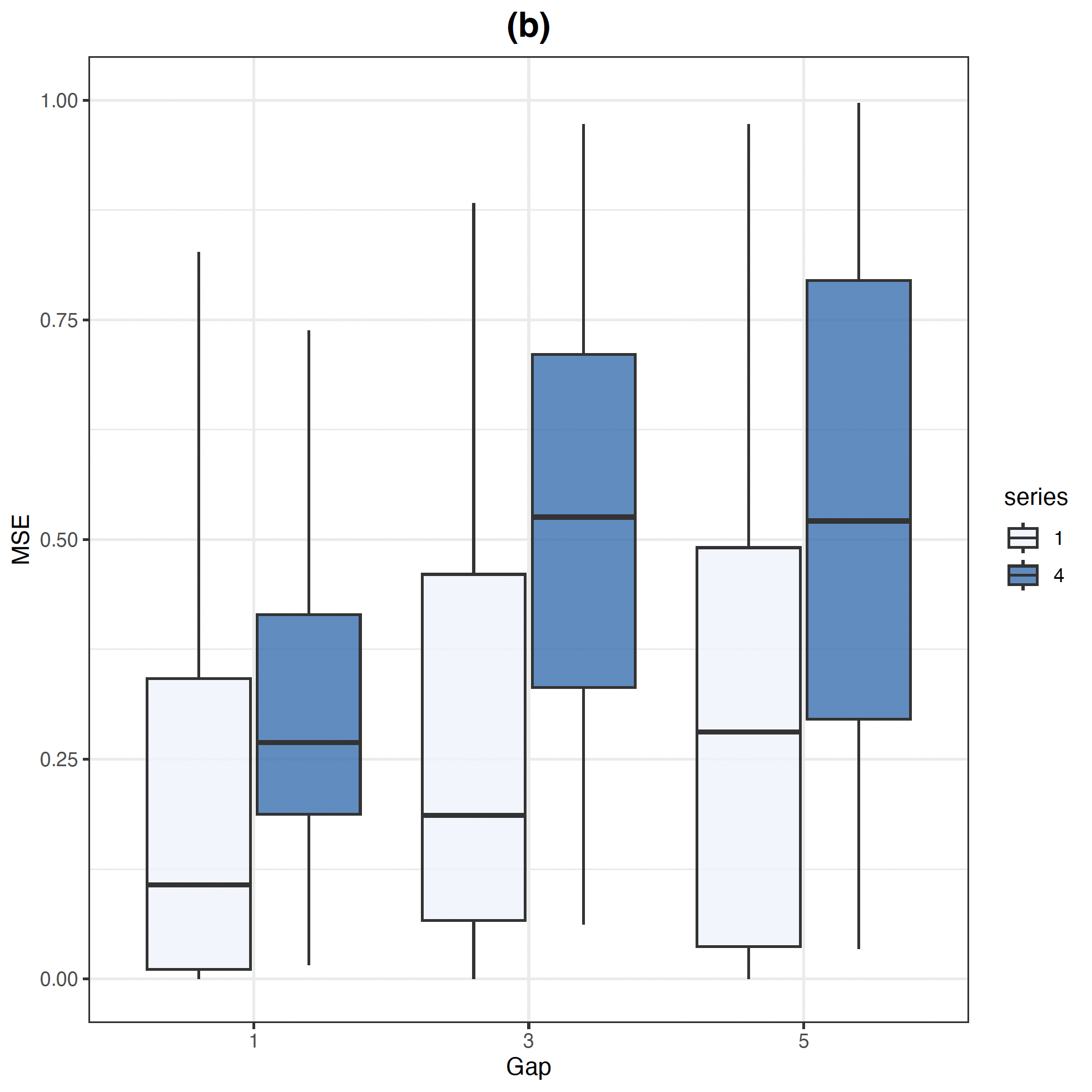}
\end{minipage}
\caption{Boxplots of the mean squared interpolation error obtained using the QiAR model on simulated irregularly observed time series with sample size $N=300$ under three different average interpolation gaps. $series$ denotes the number of time series that were imputed simultaneously. $Gap$ represents the average temporal distance between the imputed value and its previous and next observations. Figure (a) corresponds to the parameter configuration $\phi^{(0)}=0.8$, $\phi^{(1)}=0.4$, $\phi^{(2)}=0.2$, and $\phi^{(3)}=0.2$, whereas Figure (b) corresponds to the configuration $\phi^{(0)}=0.5$, $\phi^{(1)}=0.3$, $\phi^{(2)}=0.3$, and $\phi^{(3)}=0.2$. \label{fig:interpolation}}
\end{figure*}
\end{center}

Missing values were imputed by maximum likelihood, following the
approach in [15]: we first estimated the QiAR parameters, then used them
to compute the imputations. For each scenario we computed the MSE
between imputed and true values; Figure~\ref{fig:interpolation} shows
the resulting distributions.

Three patterns emerge. First, MSE increases with the gap between
observations --- closer observations interpolate more accurately.
Second, the stronger-autocorrelation configuration ($\phi^{(0)}=0.8$,
Fig.~\ref{fig:interpolation}a) yields lower MSE than the weaker one
($\phi^{(0)}=0.5$, Fig.~\ref{fig:interpolation}b). Third, MSE rises
substantially when four series are imputed simultaneously rather than
one, since the method draws on past, future, and concurrent
observations from the other series. With only one series missing,
concurrent values from the rest aid imputation; with all four missing
at once, that information is unavailable.

\subsection{Model implementation for remote sensing}

\begin{figure}   
  \centering                                                 
  \includegraphics[width=\linewidth]{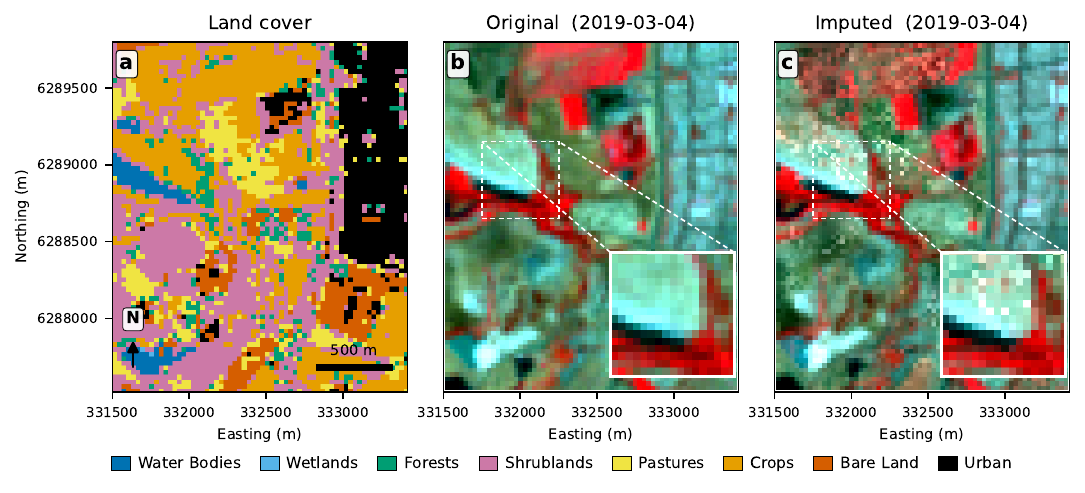}
  \caption{Remote sensing dataset and scene reconstruction.         
\textbf{(a)} Land cover classification of the study area              
($\sim$440~ha, 4{,}864 pixels at 30~m resolution), spanning seven     
cover types \cite{ZHAO2016170}. \textbf{(b)} False-colour             
NIR--Red--Green composite of the original Landsat/Sentinel-2
reflectance for 4~March 2019, where healthy vegetation appears in red.
\textbf{(c)} Same composite after temporal gap-filling. Insets show a
2$\times$ zoom over a common high-contrast feature, highlighting how
fine-scale spatial structure is preserved, while previously missing
observations are recovered.}
  \label{fig:studyAreaRS}
\end{figure}

As a proof of concept for the Earth Observation (EO) dataset, we used time series data from Landsat 5, 7, 8, and 9 collected over 440 ha near Santiago, Chile. This account of 4,864 time series includes seven land cover types (Figure \ref{fig:studyAreaRS}a), including crops, water bodies, urban areas, and several natural vegetation types. 
We used the Landsat constellation, including Level 2 Collection 2 Tier 1 scenes from TM, ETM+, and OLI/TIRS sensors for Landsat 5, 7, 8, and 9 \cite{usgs2022landsat}. We used scenes from 2000-01-01 to 2023-12-31. We merged the collections using cloud and shadow masks based on QA\_PIXEL band information from the CFMASK algorithm \cite{foga2017cloud}.
We plotted the Normalized Difference Vegetation Index (NDVI) \cite{rouse1973paper} of the image collections using the red and near infrared (NIR) bands of the images (i.e., $NIR-red/NIR+red$). See \cite{Fuentes_2024} for a full description of the dataset.

Using these data, we ensure a sufficient number of time series for testing the proposed algorithms, while maximizing the variability of time series behavior by including different land cover and vegetation types with distinct phenology \cite{Fuentes_2024,Fuentes2025-sf}. As a showcase, we selected four of the most commonly used reflectance bands for the multivariate analysis: blue, green, red, and near-infrared.

The first implementation of the  QiAR model in this data consists on  reconstructing an image over a period of time. This exercise would be useful when the image could not be captured, a common occurrence would occur when  clouds obstruct the view between the satellite and the terrain under study.
From a temporal perspective, it is common to observe that the terrain remains constant over time. Therefore, if an image cannot be obtained on a given day, it should not differ significantly from the previous day's image or the following day relative to the unobserved day.
Therefore, interpolating the terrain image for a day on which it could not be observed can help complement the knowledge about the terrain under study. The goal is to prevent this from resulting in a loss of information, which is critical, for example, for identifying deforestation, monitoring drought evolution, and analyzing other natural phenomena.
The objective is to simultaneously fit the QiAR model to a four-dimensional time series vector composed of  the four bands (blue, green, red, and near infrared) to reconstruct missing imagery for dates with no observations. The reconstruction methodology involves a two-stage process for each pixel:
\begin{itemize}
\item Trend and Seasonality Removal: A multiple linear regression model is fitted to the historical time series of each spectral band. An algorithmic procedure generates a set of explanatory variables based on observation time (e.g., linear trend, sinusoidal components for seasonality) to model and remove these temporal components. The optimal model for each band is selected based on its ability to minimize the Mean Squared Error (MSE).
\item QiAR Modeling and Interpolation: The QiAR model is then applied to the residuals obtained from the best-fitting models for the four bands. This model simultaneously captures the underlying autocorrelation structure and the cross-correlation between the bands. Finally, this fitted QiAR model interpolates the missing values for the target dates, effectively reconstructing the unobserved imagery.
\end{itemize}


Figure \ref{fig:studyAreaRS}b,c presents the image reconstruction results. 
Significant noise is observable in both reconstructions, particularly in areas containing vegetation. Conversely, other zones, such as the right section of the terrain, were captured accurately.
Consequently, an additional experiment is conducted to quantify the reconstruction error (noise) relative to the observed land cover class. For this purpose, five pixels at a specific time were selected from each of the eight land cover classes (except Wetlands, for which only two pixels exist in total). A train-test validation procedure was employed for each pixel: the observation was withheld, the model was trained on the remaining data, and an attempt was made to predict the value of the hidden observation.
Previous work has indicated that land cover classes associated with agriculture (Crops) present greater difficulty in predicting image values. This finding was considered to verify whether the same results are obtained using the final optimized model. The measurement was based on comparing the observed value at the specific time and the value predicted by the model. Subsequently, the Mean Squared Error (MSE) was calculated for each land cover class by averaging the results from its five selected pixels.
To complement this analysis, the same experiment was slightly modified: time points adjacent to the validation observation will also be withheld. Specifically, a target date for which data is also available three weeks before and three weeks after was chosen. 

Estimations will then be made for this target date while systematically withholding various combinations of the preceding and subsequent observations. The date ``January 1, 2008'' meets this requirement in the database and was used for this experiment.
The results indicate that, across all four spectral bands, the land cover classes with the lowest prediction error are ``Pastures'' and ``Bare Land'' whereas ``Wetlands'' and ``Crops'' exhibit the highest prediction error. Notably, the ``Crops'' class is associated with agricultural activity, which confirms the earlier finding reported in previous work.
Regarding the validation results based on the presence or absence of temporally adjacent observations, the analysis fails to establish a clear relationship between improved prediction accuracy and the availability of data points closer to the target date. No conclusive evidence was found to suggest that more temporally proximal data enhances prediction for a specific time.

\subsection{Application in astronomical data}
\label{sec:astro}

Another area where the QiAR model can be useful is in astronomical data. In this field, it is common to find unequally spaced multivariate time series. The reason for the multivariate structure is that the brightness of astronomical objects is typically observed in more than one passband. For this application, we used light curve data from the Zwicky Transient Facility (ZTF) processed by the Chilean broker \textit{ALeRCE}. In ZTF, each astronomical object is observed in two optical bands ($g$ and $r$). Thus, the modeling problem is inherently bivariate, which could be handled using a BiAR model.\\

However, the QiAR model can also be applied in this context by considering that two of the four series in the process are observed and the remaining two are latent. In this way, the QiAR model provides a generalization of the BiAR model, offering greater flexibility in representing the cross-correlations between the series. This flexibility is controlled through the parameters $\alpha$ and $\beta$, as well as by defining which components of the QiAR process are treated as observable. Note that this allows for six different configurations of the QiAR model that can be adjusted for each example.\\

In the following, several examples are presented in which the QiAR model is fitted to astronomical objects. It is important to note that the observational times in the two bands are not identical, although in many cases they differ only by a small gap. To address this, we use a time-matching procedure, which assumes that two observations correspond to the same time when their timestamps differ by less than a small tolerance. If no match is found, the missing value in the other series is imputed. This procedure is described in more detail in \cite{Elorrieta_2021}. After the matching step, both light curves are detrended using a loess model fitted separately to each band. Finally, all six QiAR configurations are fitted, and the best configuration is selected as the one that minimizes the root mean squared error (RMSE).\\

The first example corresponds to a light curve of quasi-stellar object (QSO) identified as ``ZTF18abxfcqm''. The matched light curve for this object has 172 observations measured over a range of 1799 days. In this case, the best-fitting model corresponds to the QiAR configuration in which the first and fourth components are observable, while the remaining components are latent. Furthermore, the estimated parameters $\alpha$ and $\beta$ are equal to one, meaning that the fitted model corresponds to the standard QiAR specification. For the remaining parameters of the QiAR process, the estimates were $\hat{\phi}^{(0)} = 0.991$, $\hat{\phi}^{(1)} = 0.007$,  $\hat{\phi}^{(2)} = 0.012$, and $\hat{\phi}^{(3)} = 0.007$.
Based on this result, the best-fitting QiAR model can be represented by the following state space system:

\begin{eqnarray}
\left( \begin{array}{c}
y_{t_{j}}^{(0)}\\
y_{t_{j}}^{(1)}\\
y_{t_{j}}^{(2)}\\
y_{t_{j}}^{(3)}
\end{array}\right) &=&F_{t_j} \left( \begin{array}{c}
y_{t_{j-1}}^{(0)}\\
y_{t_{j-1}}^{(1)}\\
y_{t_{j-1}}^{(2)}\\
y_{t_{j-1}}^{(3)}
\end{array}\right)+V_{t_j}\\
\left( \begin{array}{c}
y_{t_{j}}^{(0)}\\
y_{t_{j}}^{(2)}\\
\end{array}\right) &=& \left(\begin{array}{cccc} 1 & 0 & 0 & 0 \\
0 & 0 & 0 & 1 \\
\end{array} \right) \left( \begin{array}{c}
y_{t_{j}}^{(0)}\\
y_{t_{j}}^{(1)}\\
y_{t_{j}}^{(2)}\\
y_{t_{j}}^{(3)}
\end{array}\right)+W_{t_j}, \quad j=1,\ldots,n.
\end{eqnarray}

where the transition matrix \( F_{t_j} \) for a unit time gap is defined as:

\begin{equation}
F_{t_j} = \left(\begin{array}{cccc} 0.991 & -0.007 & -0.012 & -0.007 \\
0.007 & 0.991 & -0.007 & 0.012 \\
0.012 & 0.007 & 0.991 & -0.007 \\
0.007 & -0.012 & 0.007 & 0.991
\end{array} \right)
\end{equation}

and the observation matrix \( G_{t_j} \) is given by:

\begin{equation}
G_{t_j} = \left(\begin{array}{cccc} 1 & 0 & 0 & 0 \\
0 & 0 & 0 & 1 \\
\end{array} \right)
\end{equation}

\begin{center}
\begin{figure*}
\begin{minipage}{0.32\linewidth}
\centering
\includegraphics[width=\textwidth]{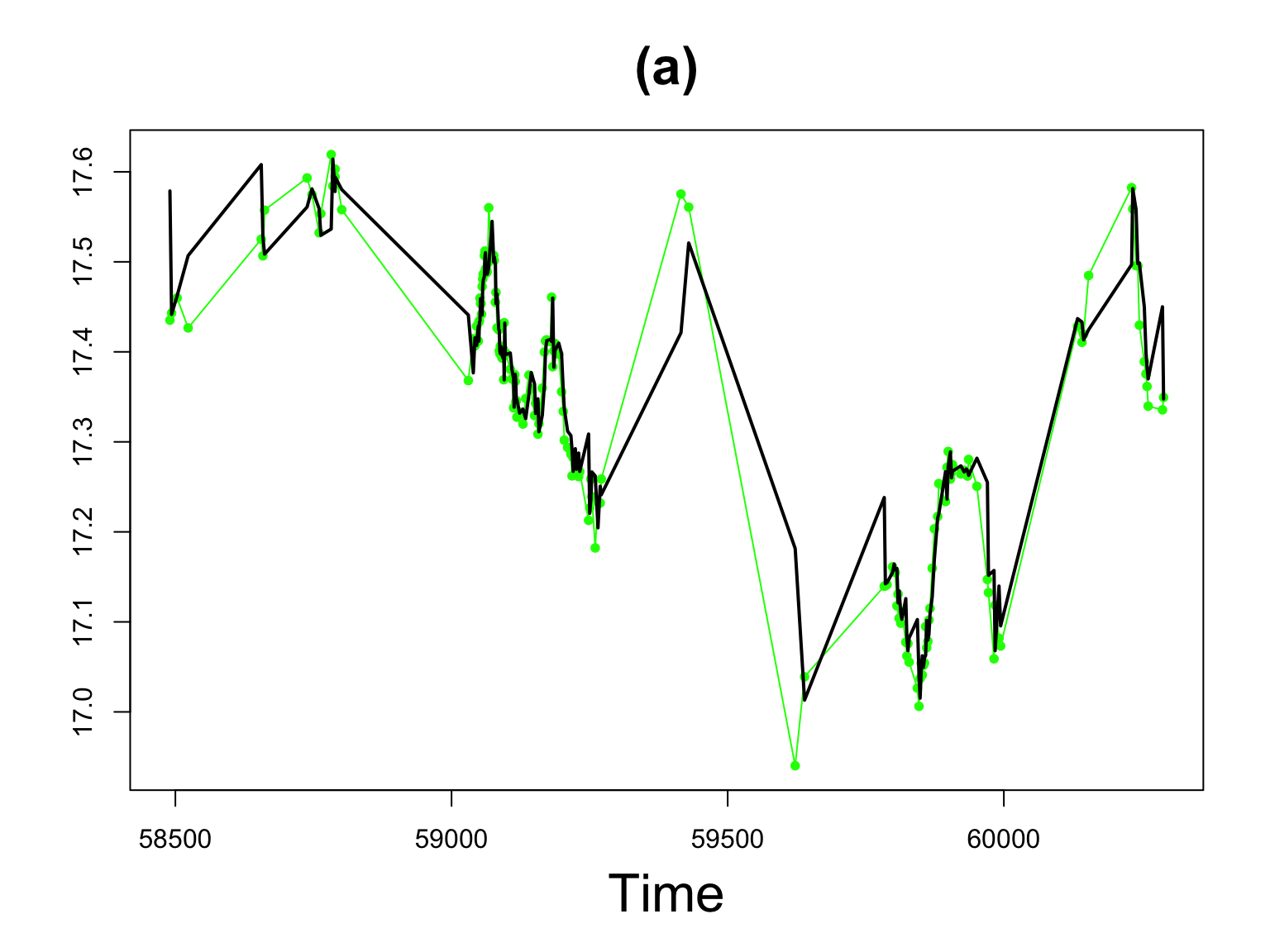}
\end{minipage}
\begin{minipage}{0.32\linewidth}
\centering
\includegraphics[width=\textwidth]{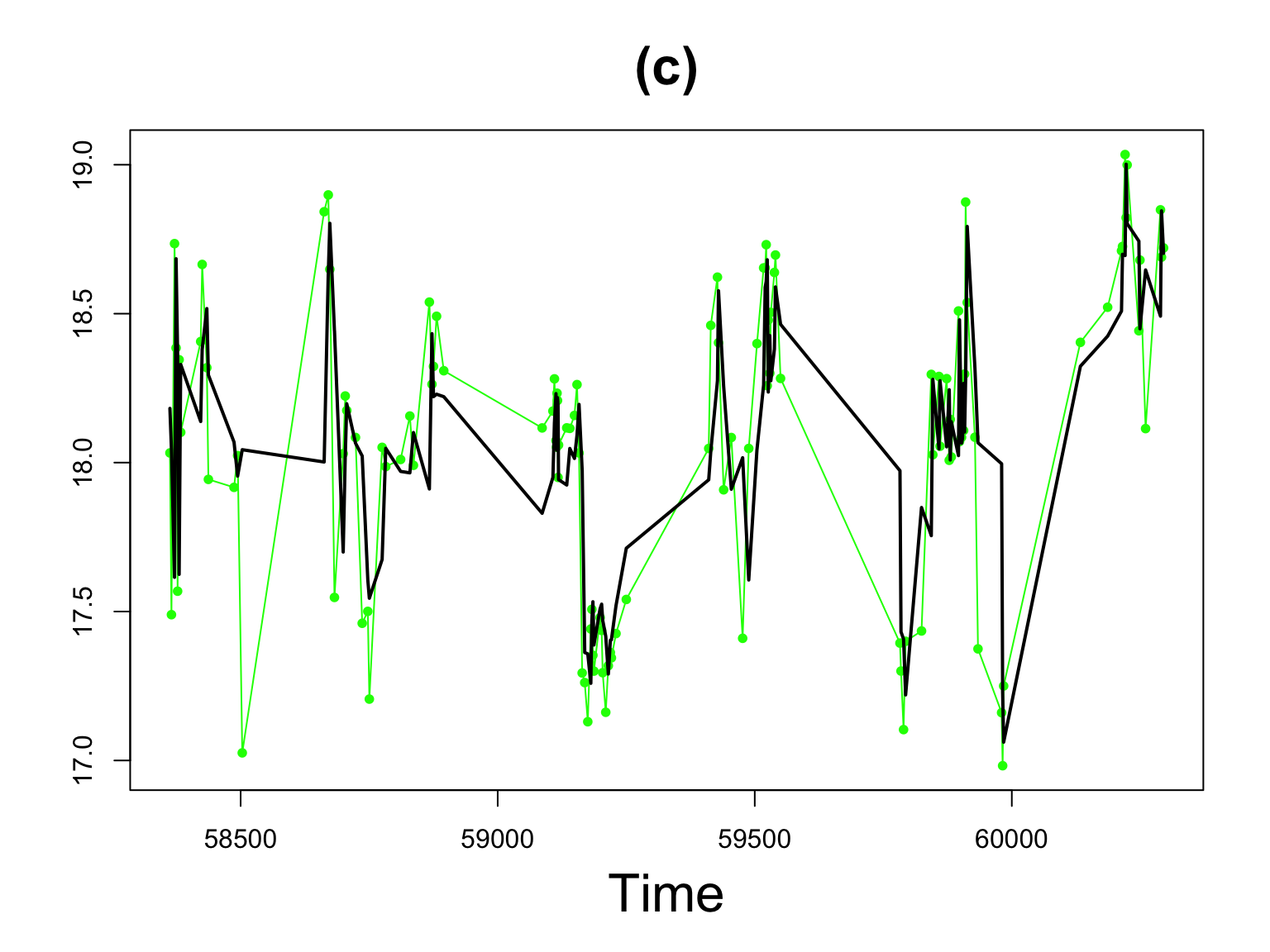}
\end{minipage}
\begin{minipage}{0.32\linewidth}
\centering
\includegraphics[width=\textwidth]{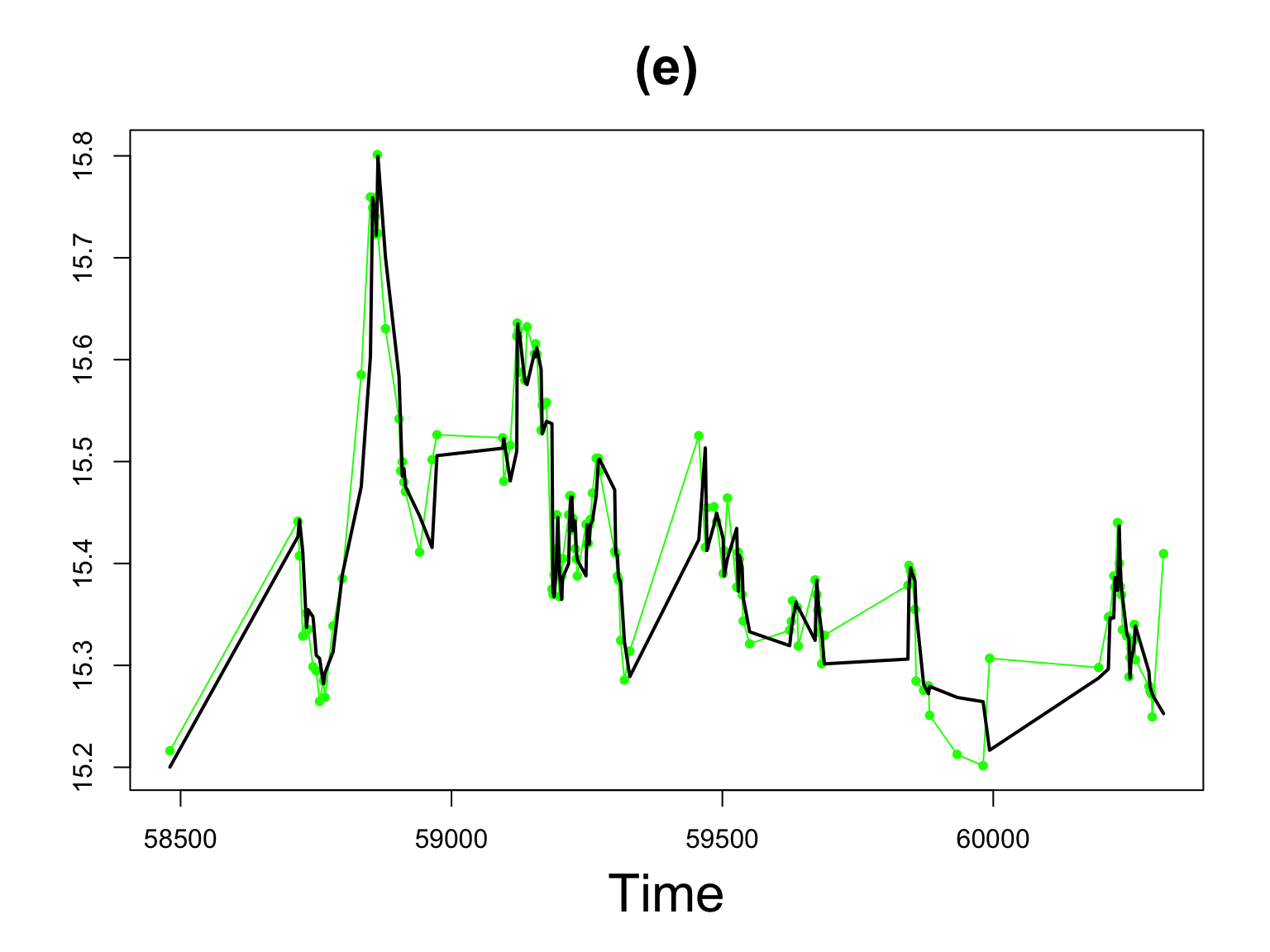}
\end{minipage}
\begin{minipage}{0.32\linewidth}
\centering
\includegraphics[width=\textwidth]{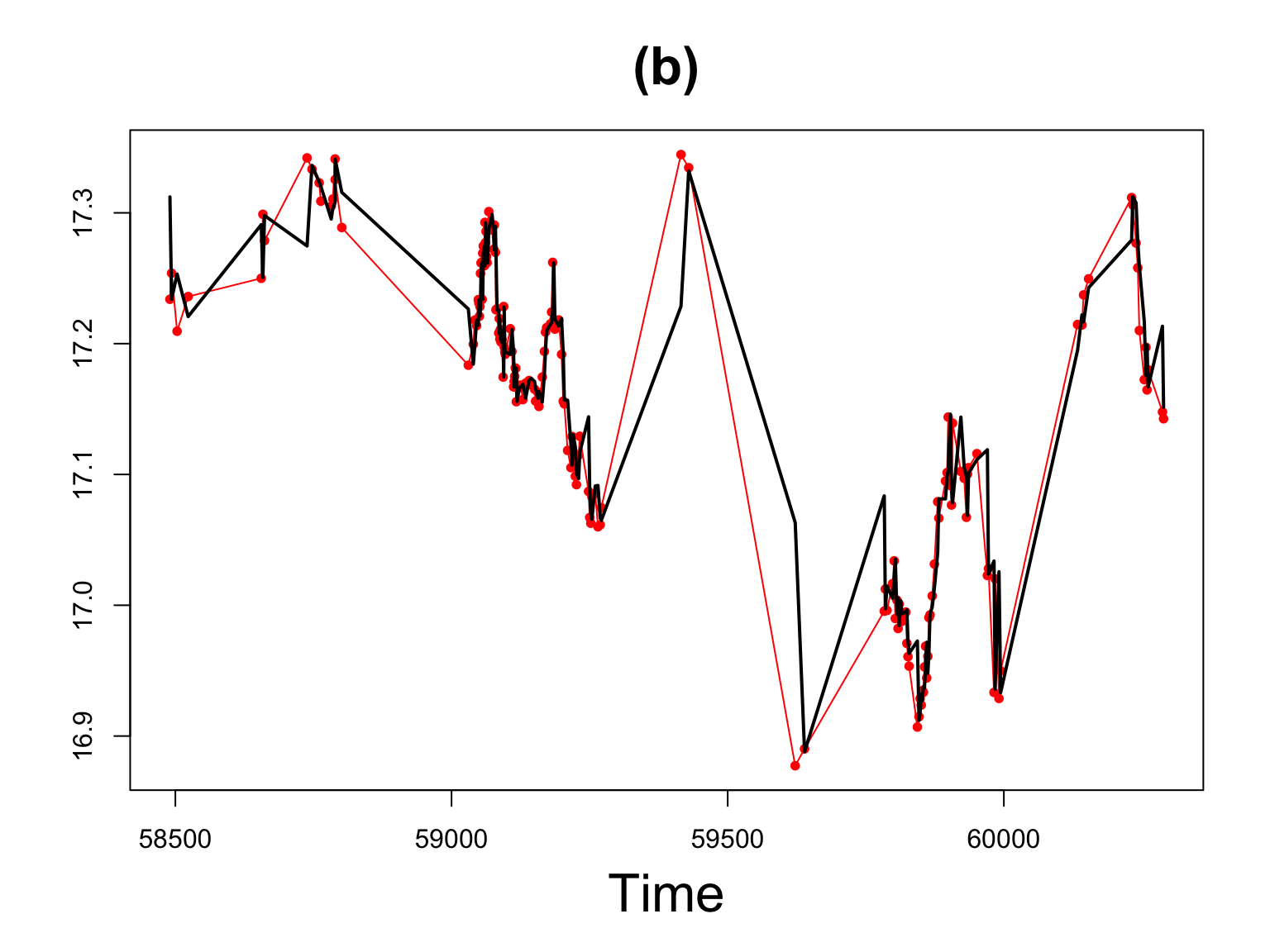}
\end{minipage}
\begin{minipage}{0.32\linewidth}
\centering
\includegraphics[width=\textwidth]{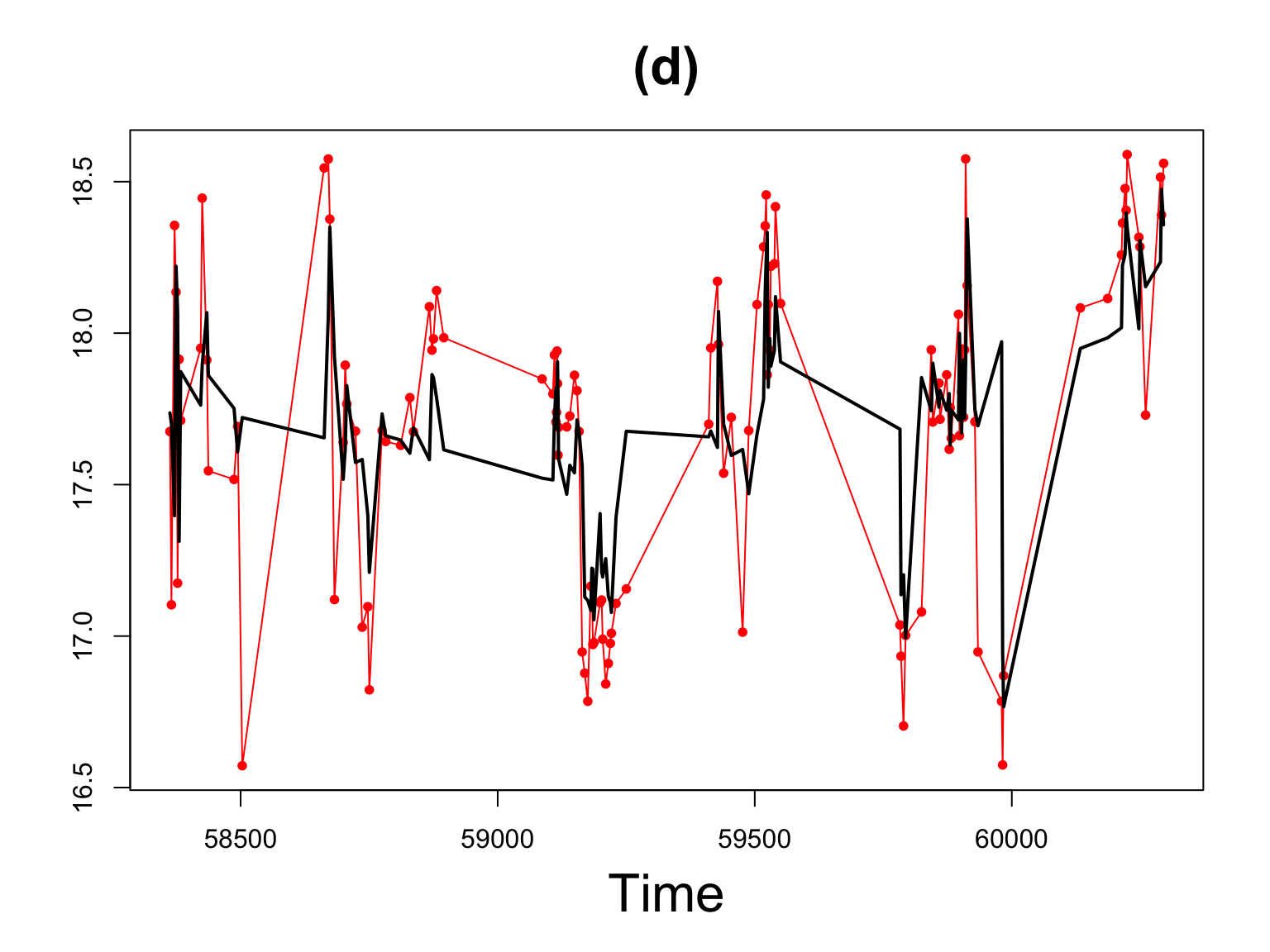}
\end{minipage}
\begin{minipage}{0.32\linewidth}
\centering
\includegraphics[width=\textwidth]{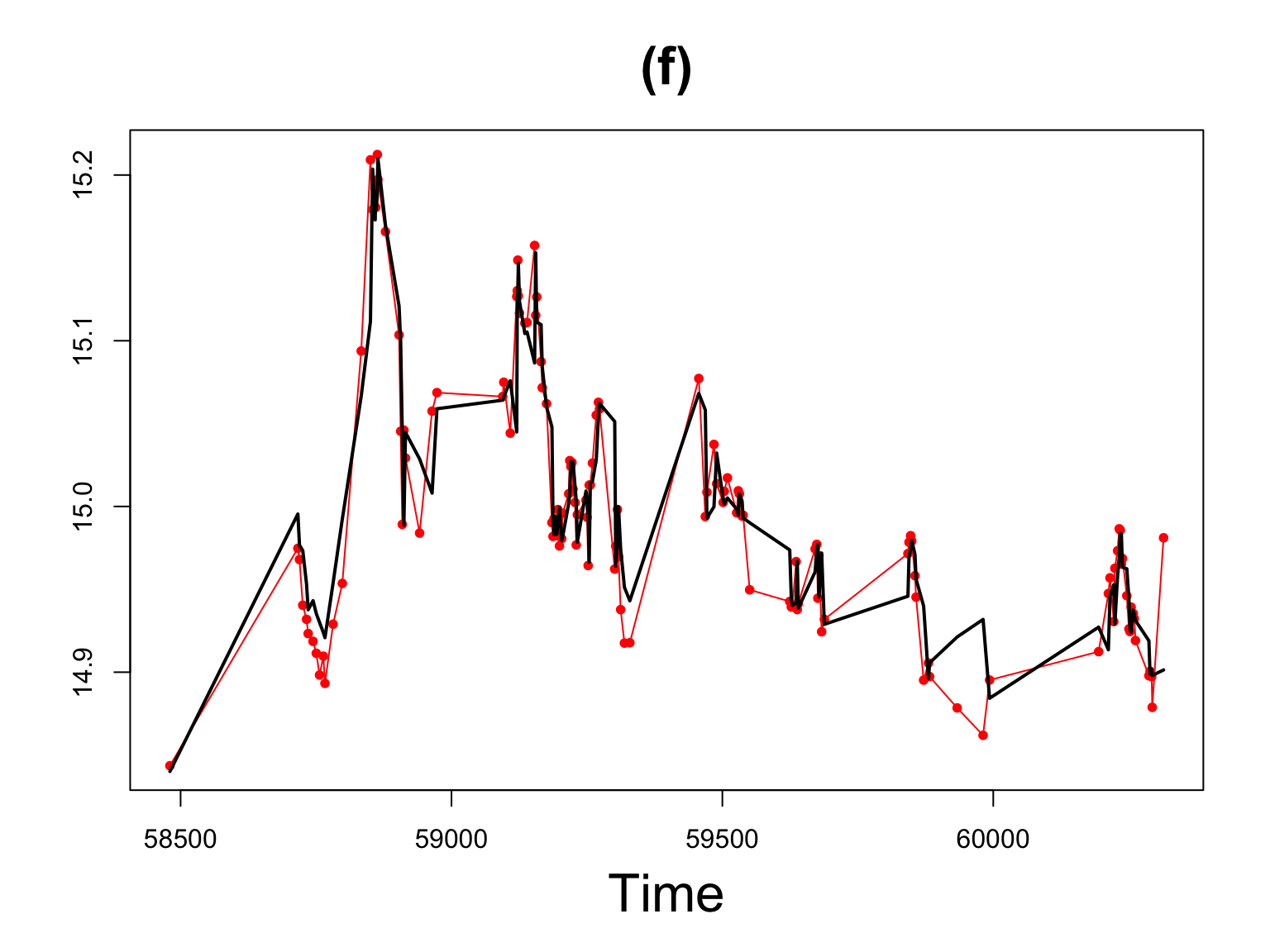}
\end{minipage}
\caption{Panels (a)--(b): QiAR model fit for the QSO object \textit{ZTF18abxfcqm}. Panels (c)--(d): QiAR model fit for the blazar \textit{ZTF18abosoyb}. Panels (e)--(f): QiAR model fit for the AGN \textit{ZTF18aaaexud}. The plots show the observed (green for $g$ band and red for $r$ band) and fitted (black line) light curves for each object. \label{fig:astro}}
\end{figure*}
\end{center}

The best fitted model achieved a root mean square error of $0.2836$. Considering that the data has a standard deviation of $0.8058$, this implies that the QiAR model reduces variability by 64.81\%. Additionally, the Box–Ljung test statistics at lag 15 for the residuals in the $g$ and $r$ bands are $BL(15)_g = 0.896$ and $BL(15)_r = 0.117$, respectively, indicating white noise residuals. In comparison, a VAR(1) model yields an RMSE of $0.390$ with Box--Ljung p-values of $BL(15)_g = 0.051$ and $BL(15)_r = 0.019$, suggesting remaining autocorrelation in the residuals. Overall, these results indicate that the QiAR model provides a better fit to the data, as illustrated in Figure~\ref{fig:astro} a)-b), producing uncorrelated residuals in both bands and effectively capturing the temporal dependency structure. \\

The second example corresponds to a Blazar object identified as ``ZTF18abosoyb''. This light curve has 123 observations measured over a range of 1933 days. In this case, the estimated parameters are $\hat{\phi}^{(0)} = 0.967$, $\hat{\phi}^{(1)} = 0.003$, $\hat{\phi}^{(2)} = -0.022$, $\hat{\phi}^{(3)} = -0.054$, $\alpha = 0$, and $\beta = 1$. Particularly, the value $\alpha = 0$ implies that some elements of the transition matrix are equal to zero, as can be seen in the following transition matrix of the fitted model:

\begin{equation}
F_{t_j} = \left(\begin{array}{cccc} 0.967 & 0 & 0.021 & 0 \\
0.003 & 0.967 & 0.054 & -0.021 \\
-0.021 & 0 & 0.967 & 0 \\
-0.054 & 0.021 & 0.003 & 0.967 \\
\end{array} \right)
\end{equation}

Similar to the previous case, the best configuration assumes the first and fourth elements of the model as observable and the remaining two as latent, resulting in the same observation matrix. Under this specification, the model achieves an RMSE of $0.748$, which is slightly lower than that obtained with a VAR(1) model ($0.796$). For this object, both models produce white-noise residuals, with Box--Ljung $p$-values for both bands well above the 0.05 significance threshold. The statistical diagnostics confirm that the QiAR model provides an adequate fit to the data, as illustrated in Figure~\ref{fig:astro}c)--d).\\

The last example corresponds to the light curve of an AGN object identified as ``ZTF18aaaexud'' with 139 observations spanning 1834 days. The parameters estimated for the QiAR model are $\hat{\phi}^{(0)} = 0.992$, $\hat{\phi}^{(1)} = 0.018$, $\hat{\phi}^{(2)} = -0.011$, $\hat{\phi}^{(3)} = -0.006$, $\alpha = 1$, and $\beta = 0.5$. The transition matrix for a unit time gap of the model fitted is as follows:

\begin{equation}
F_{t_j} =
\left(
\begin{array}{cccc}
0.992 & -0.018 & 0.005 & 0.003 \\
0.018 & 0.992 & 0.003 & -0.005 \\
-0.011 & -0.006 & 0.992 & -0.018 \\
-0.006 & 0.011 & 0.018 & 0.992
\end{array}
\right)
\end{equation}

The configuration yielding the best fit corresponds to the model in which the first and third inputs are observed, while the remaining components are latent, as shown in the following observation matrix \( G_{t_j} \):

\begin{equation}
G_{t_j} = \left(\begin{array}{cccc} 1 & 0 & 0 & 0 \\
0 & 0 & 1 & 0 \\
\end{array} \right)
\end{equation}

The QiAR model provides an accurate fit to the light curve, as reflected in both the fitted values and the whiteness of the residuals. The model achieves an RMSE of $0.301$, corresponding to a 62.21\% reduction in residual variability relative to the original series standard deviation. Moreover, the Box-Ljung tests yield $p$-values greater than 0.05 for both the $g$ and $r$ bands, indicating no significant autocorrelation in the residuals. In contrast, the VAR(1) model produces a $p$-value of $0.031$ for the $g$-band residuals and a higher RMSE of $0.343$, confirming that the QiAR model outperforms the VAR(1) for this AGN light curve.\\

\section{Discussion}

This study introduces a new framework for modeling multivariate time series observed at irregular time intervals. Unlike classical autoregressive models that rely on regularly spaced observations, the proposed approach incorporates the temporal structure of irregular observation gaps through a hypercomplex representation. By embedding multivariate observations into a hypercomplex algebraic structure, the proposed methodology provides a compact representation capable of capturing cross-variable interactions and temporal dynamics simultaneously.
An important feature of the proposed framework is that the hypercomplex formulation admits a structured matrix representation, allowing the model to be expressed in terms of a state-space system. This representation bridges the algebraic formulation with well-established tools from time-series analysis and signal processing.
The state-space representation enables parameter estimation and inference using Kalman filtering techniques. In this formulation, the hypercomplex autoregressive process defines the evolution equation of the latent state, while the observed multivariate series correspond to the measurement equation. The Kalman filter provides an efficient recursive algorithm for estimating the hidden states and updating parameter estimates as new observations become available.
This estimation framework is particularly well suited for irregularly sampled data, as the time-gap structure can be naturally incorporated into the state transition dynamics. As a result, the proposed model allows both prediction and parameter estimation to be performed in a computationally efficient manner.
Irregular sampling is common in many scientific disciplines. In astronomy, observational schedules depend on telescope availability, weather conditions, and orbital constraints. Similarly, remote sensing data are often affected by satellite revisit cycles, cloud coverage, and sensor limitations. These factors frequently produce multivariate time series with nonuniform observation gaps that challenge traditional modeling approaches.
The proposed framework directly addresses this issue by combining hypercomplex representations with a state-space formulation that allows the irregular temporal structure to be explicitly modeled and estimated using Kalman filtering methods.
The hypercomplex autoregressive formulation offers several advantages. First, it provides a unified representation for multivariate dynamics within a single algebraic structure. Second, the framework naturally accommodates irregular observation gaps without requiring interpolation or resampling of the data. Third, the state-space representation allows the model to be estimated using established recursive filtering techniques, making it computationally feasible for practical applications.
While this work focuses on applications to astronomy and remote sensing, the proposed framework has broad applicability across scientific disciplines that routinely produce irregularly sampled multivariate time series. Examples include neuroscience, where neural recordings from multiple electrodes are collected at varying rates; epidemiology, where disease surveillance data arrive at nonuniform time intervals across multiple health indicators; finance, where asset prices and trading volumes are observed asynchronously across markets; and climate science, where proxy records such as ice cores and sediment layers provide multivariate measurements at irregular temporal resolutions. In each of these domains, the ability to model cross-variable dependencies under irregular sampling — without requiring interpolation or resampling — represents a meaningful methodological advance over existing approaches.
Beyond interpolation and forecasting, the state-space formulation of the proposed model opens natural avenues for change point detection and structural break analysis. In astronomy, the identification of abrupt changes in the autocorrelation or cross-correlation structure of multi-band light curves could signal novel physical phenomena, such as state transitions in active galactic nuclei, precursor activity in variable stars, or the onset of transient events. The Kalman filter residuals, combined with the model's parsimonious parameterization, provide a principled basis for detecting such anomalies in streaming time-domain data. In remote sensing, the same methodology can be applied to identify changes in the environment from multi-spectral satellite observations — including the detection of deforestation, land degradation, and shifts in vegetation phenology driven by drought or other climate-related disturbances. By jointly modeling multiple spectral bands as an irregularly observed multivariate process, the framework can distinguish genuine structural change from noise or seasonal variation, thereby supporting more reliable environmental monitoring at scale.

\newpage
\section{Appendix}

\subsection{ Proof of Theorem 1}

 The proof of this  theorem is based on the Euler's and the de Moivre's formulas for complex, quaternions and octonions random variables. Let $\delta_{s}=t_s-t_{s-1}$ and assume first that $\bm{\phi}=\phi_0+i\phi_1$ is a complex parameter. It has a polar representation as $\bm{\phi}=\|\bm{\phi}\|(\cos(\theta)+i\sin(\theta))$, where $\|\bm{\phi}\|^2=\phi_0^2+\phi_1^2$, and $\theta=\tan^{-1}(\frac{\phi_1}{\phi_0})$. Moreover, by the De Moivre's formula, for any rational number $\delta_{s}$, $\bm{\phi}^{\delta_{s}}$ can be expressed as $\bm{\phi}^{\delta_{s}}=\|\bm{\phi}\|^{\delta_{s}}(\cos(\delta_{s}\theta)+i\sin(\delta_{s}\theta))$. Therefore, the process  described by Eq. (\ref{eq:its}) can be expresses as $\bm{y}_{t_js}=\bm{\alpha}_{t_js}\bm{y}_{t_{js-1}} +\bm{\sigma_{t_{s}}}$ as required, where ${\bm{\alpha}}_{t_s}={\alpha}_{t_s}^{(0)}+i{\alpha}_{t_s}^{(1)}$ with ${\alpha}_{t_s}^{(0)}=\|\bm{\phi}\|^{\delta_{s}}\cos(\delta_{s}\theta)$ and ${\alpha}_{t_s}^{(1)}=\|\bm{\phi}\|^{\delta_{s}}\sin(\delta_{s}\theta)$.


Assume now a model on quaternion random variables.\\

$x_{t_s} = \phi^{\delta_{s}} x_{t_{s-1}} + \sigma_{t_s}\varepsilon_{t_s}$. We are interested in expressing the term $\phi^{\delta_{s}} = \alpha_{t_s}$. Note that for $\|\phi\|^2=\phi^{(0)2}+\phi^{(1)2} + \phi^{(2)2} +\phi^{(3)2}$

\begin{eqnarray*}
\phi^{t_s-t_{s-1}} &=&  (\phi^{(0)}+\phi^{(1)} i + \phi^{(2)} j +\phi^{(3)} k)^{\delta_{s}} = \|\phi\|^{\delta_{s}} \left(\frac{\phi^{(0)}+\phi^{(1)} i + \phi^{(2)} j +\phi^{(3)} k}{\|\phi\|} \right)^{\delta_{s}}
\end{eqnarray*}

\noindent
Using the polar representation of quaternions we obtain,

\begin{eqnarray*}
\left(\frac{\phi^{(0)}+\phi^{(1)} i + \phi^{(2)} j +\phi^{(3)} k}{\|\phi\|} \right)^{\delta_{s}} &=& (\cos(\psi) + u \sin(\psi))^{\delta_{s}}\\
\left(\phi^{(0)}+\phi^{(1)} i + \phi^{(2)} j +\phi^{(3)} k \right) ^{\delta_{s}}&=& \|\phi\|^{\delta_{s}} (\cos(\psi) + u \sin(\psi))^{\delta_{s}}
\end{eqnarray*}

\noindent
where $u = \frac{\phi^{(1)} i + \phi^{(2)} j +\phi^{(3)} k}{\sqrt{\phi^{(1)2} + \phi^{(2)2} + \phi^{(3)2}}}$ and $\psi=\arccos{(\frac{\phi^{(0)}}{\|\phi\|})}$. By the De Moivre's formula we get,

\begin{eqnarray*}
\|\phi\|^{\delta_{s}} (\cos(\psi) + u \sin(\psi))^{\delta_{s}} &=& \|\phi\|^{\delta_{s}} (\cos(\delta_{s}\psi) + u \sin(\delta_{s}\psi))\\ 
&=& \|\phi\|^{\delta_{s}} \cos(\delta_{s}\psi) + \frac{\|\phi\|^{\delta_{s}}}{\sqrt{\phi^{(1)2} + \phi^{(2)2} + \phi^{(3)2}}} (\phi^{(1)} \sin(\delta_{s}\psi) i  + \\ 
&&\phi^{(2)}  \sin(\delta_{s}\psi) j +\phi^{(3)}  \sin(\delta_{s}\psi) k)\\
&=& \bm{\alpha_{t_s}}  \\
\end{eqnarray*}

\noindent
where ${\alpha}_{t_s}^{(0)}= \|\phi\|^{\delta_{s}} \cos(\delta_{s}\psi)$ and ${\alpha}_{t_s}^{(l)}= \frac{\|\phi\|^{\delta_{s}}}{\sqrt{\phi^{(1)2} + \phi^{(2)2} + \phi^{(3)2}}} \phi^{(l)} \sin(\delta_{s}\psi)$ for $l=1,2,3$.

The proof follows similarly for the octonions. Using the polar representation for octonions, $\bm{\phi}^{\delta_s}$ can be expressed as

\begin{eqnarray*}
\left(\sum_{l=0}^7\phi^{(l)}f_l\right) ^{\delta_{s}}&=& \|\bm{\phi}\|^{\delta_{s}} (\cos(\psi) + w \sin(\psi))^{\delta_{s}}
\end{eqnarray*}

\noindent
where $w=\frac{\sum_{l=1}^7\phi^{(l)}f_l}{\sqrt{\sum_{l=1}^7\phi^{(l)2}}}$. Using the de Moivre's formula

\begin{eqnarray*}
\left(\sum_{l=0}^7\phi^{(l)}f_l\right) ^{\delta_{s}}&=& \|\bm{\phi}\|^{\delta_{s}} (\cos(\delta_{s}\psi) + w \sin(\delta_{s}\psi))=\bm{\alpha_{t_s}}
\end{eqnarray*}

\noindent
where ${\alpha}_{t_s}^{(0)}= \|\phi\|^{\delta_{s}} \cos(\delta_{s}\psi)$ and ${\alpha}_{t_s}^{(l)}= \frac{\|\phi\|^{\delta_{s}}}{\sqrt{\sum_{l=1}^7\phi^{(l)2}}} \phi^{(l)} \sin(\delta_{s}\psi)$ for $l=1,\ldots,7$. 


\subsection{ Proof of Theorem 2}

To prove Theorem 2, we use the matrix representation of the multiplication operation. For the complex, in Eq (\ref{eq:itsMoivre}) the complex multiplication between $\bm{\alpha_{t_2}}$ and $\bm{y_{t_{s-1}}}$ is equal to

\[\bm{\alpha_{t_s}}\bm{y}_{t_{s-1}}=(\alpha^{(0)}_{t_s}y_{t_{s-1}}^{(0)}-\alpha^{(1)}_{t_s}y_{t_{s-1}}^{(1)})+i(\alpha^{(0)}_{t_s}y_{t_{s-1}}^{(1)}+\alpha^{(1)}_{t_s}y_{t_{s-1}}^{(0)})\]

\noindent
which can be represented in matrix notation

\[\bm{\alpha_{t_s}}\bm{y}_{t_{s-1}}=\left(\begin{array}{cc} 
\alpha^{(0)}_{t_s} & -\alpha^{(1)}_{t_s}\\\alpha^{(1)}_{t_s}&\alpha^{(0)}_{t_s}\end{array}
\right)\left( \begin{array}{c}
y_{t_{s-1}}^{(0)}\\
y_{t_{s-1}}^{(1)}
\end{array}\right).\]

This enables a representation in matrix notation of a two dimensional model as follows

\[\left( \begin{array}{c}
y_{t_{s}}^{(0)}\\
y_{t_{s}}^{(1)}
\end{array}\right)=\left(\begin{array}{cc} 
\alpha^{(0)}_{t_s} & -\alpha^{(1)}_{t_s}\\\alpha^{(1)}_{t_s}&\alpha^{(0)}_{t_s}\end{array}
\right)\left( \begin{array}{c}
y_{t_{s-1}}^{(0)}\\
y_{t_{s-1}}^{(1)}
\end{array}\right) + \left( \begin{array}{c}
\epsilon_{t_{s}}^{(0)}\\
\epsilon_{t_{s}}^{(1)}
\end{array}\right).\]

Denoting $F_{ts}=\left(\begin{array}{cc} 
\alpha^{(0)}_{t_s} & -\alpha^{(1)}_{t_s}\\\alpha^{(1)}_{t_s}&\alpha^{(0)}_{t_s}\end{array}
\right)$, the complex model can be represented as a real two dimensional vector model 

\begin{equation}
\bm{y}_{t_j}=F_{t_j}\bm{y}_{t_{j-1}} +\bm{\epsilon_{t_{j}}}\label{eq:sstransition}
\end{equation}

\noindent
where now $\bm{y}_{t_j}, \bm{\epsilon}_{t_{j}}$ are two dimensional random vectors.

Eq. (\ref{eq:sstransition}) can also be considered as the transition equation in the state space representation of the model.

The multiplication between $\bm{\alpha}_{t_2}$ and $\bm{y}_{t_{s-1}}$ in the quaternion case is equal to

\begin{eqnarray*}
\bm{\alpha}_{t_s}\bm{y}_{t_{s-1}}&=&(\alpha^{(0)}_{t_s}y_{t_{s-1}}^{(0)}-\alpha^{(1)}_{t_s}y_{t_{s-1}}^{(1)}-\alpha^{(2)}_{t_s}y_{t_{s-1}}^{(2)}-\alpha^{(3)}_{t_s}y_{t_{s-1}}^{(3)})\\
&&+ i(\alpha^{(0)}_{t_s}y_{t_{s-1}}^{(1)}+\alpha^{(1)}_{t_s}y_{t_{s-1}}^{(0)}+\alpha^{(2)}_{t_s}y_{t_{s-1}}^{(3)}-\alpha^{(3)}_{t_s}y_{t_{s-1}}^{(2)})\\
&&+j(\alpha^{(0)}_{t_s}y_{t_{s-1}}^{(2)}+\alpha^{(2)}_{t_s}y_{t_{s-1}}^{(0)}+\alpha^{(3)}_{t_s}y_{t_{s-1}}^{(1)}-\alpha^{(1)}_{t_s}y_{t_{s-1}}^{(3)}) \\
&&+k(\alpha^{(0)}_{t_s}y_{t_{s-1}}^{(3)}+\alpha^{(3)}_{t_s}y_{t_{s-1}}^{(0)}+\alpha^{(1)}_{t_s}y_{t_{s-1}}^{(2)}-\alpha^{(2)}_{t_s}y_{t_{s-1}}^{(1)})
\end{eqnarray*}

\noindent
which can also be represented in matrix notation as a four-dimensional model

\[ \left( \begin{array}{c}
y_{t_{s}}^{(0)}\\
y_{t_{s}}^{(1)}\\
y_{t_{s}}^{(2)}\\
y_{t_{s}}^{(3)}
\end{array}\right)=\left(\begin{array}{cccc} \alpha_{t_s}^{(0)} & -\alpha_{t_s}^{(1)}  & -\alpha_{t_s}^{(2)} & -\alpha_{t_s}^{(3)} \\ \alpha_{t_s}^{(1)} & \alpha_{t_s}^{(0)} & -\alpha_{t_s}^{(3)} & \alpha_{t_s}^{(2)} \\ \alpha_{t_s}^{(2)} & \alpha_{t_s}^{(3)} & \alpha_{t_s}^{(0)} & -\alpha_{t_s}^{(1)}  \\ \alpha_{t_s}^{(3)} & -\alpha_{t_s}^{(2)} & \alpha_{t_s}^{(1)}  & \alpha_{t_s}^{(0)} \end{array} \right)\left( \begin{array}{c}
y_{t_{s-1}}^{(0)}\\
y_{t_{s-1}}^{(1)}\\
y_{t_{s-1}}^{(2)}\\
y_{t_{s-1}}^{(3)}
\end{array}\right)+\left( \begin{array}{c}
\epsilon_{t_{s}}^{(0)}\\
\epsilon_{t_{s}}^{(1)}\\
\epsilon_{t_{s}}^{(2)}\\
\epsilon_{t_{s}}^{(3)}
\end{array}\right)\]

Similar to the complex case, by denoting
\begin{equation}
F_{t_s} = \left(\begin{array}{cccc} \alpha_{t_s}^{(0)} & -\alpha_{t_s}^{(1)}  & -\alpha_{t_s}^{(2)} & -\alpha_{t_s}^{(3)} \\ \alpha_{t_s}^{(1)} & \alpha_{t_s}^{(0)} & -\alpha_{t_s}^{(3)} & \alpha_{t_s}^{(2)} \\ \alpha_{t_s}^{(2)} & \alpha_{t_s}^{(3)} & \alpha_{t_s}^{(0)} & -\alpha_{t_s}^{(1)}  \\ \alpha_{t_s}^{(3)} & -\alpha_{t_s}^{(2)} & \alpha_{t_s}^{(1)}  & \alpha_{t_s}^{(0)} \end{array} \right)\end{equation} 

\noindent
the quaternion model can be represented as a model for four-dimensional random variables. 

\begin{equation}
\bm{y}_{t_j}=F_{t_j}\bm{y}_{t_{j-1}} +\bm{\epsilon_{t_{j}}}\label{eq:sstransitionH}
\end{equation}

And Eq. (\ref{eq:sstransitionH}) can also be considered as the transition equation in the state space representation of the model.

For the generalized quaternions, the matrix $F_{t_s}$ is the following:

\begin{equation}\label{eq:Fquatgen2}
F_{t_s} = \left(\begin{array}{cccc} \alpha_{t_s}^{(0)} & -\alpha\alpha_{t_s}^{(1)}  & -\beta\alpha_{t_s}^{(2)} & -\alpha\beta\alpha_{t_s}^{(3)} \\ 
\alpha_{t_s}^{(1)} & \alpha_{t_s}^{(0)} & -\beta\alpha_{t_s}^{(3)} & \beta\alpha_{t_s}^{(2)} \\ 
\alpha_{t_s}^{(2)} & \alpha\alpha_{t_s}^{(3)} & \alpha_{t_s}^{(0)} & -\alpha\alpha_{t_s}^{(1)}  \\ 
\alpha_{t_s}^{(3)} & -\alpha_{t_s}^{(2)} & \alpha_{t_s}^{(1)}  & \alpha_{t_s}^{(0)} \end{array} \right).\end{equation} 

As in the previous cases, the multiplication between $\bm{\alpha}_{t_s}$ and $\bm{y}_{t_{s-1}}$ when these variables are octonions, has a matrix representation  such that when $F_{t_s}$ is equal to

\begin{equation}
F_{t_s} = \left(\begin{array}{cccccccc} 
\alpha_{t_s}^{(0)} & -\alpha_{t_s}^{(1)}  & -\alpha_{t_s}^{(2)} & -\alpha_{t_s}^{(3)} & -\alpha_{t_s}^{(4)} & -\alpha_{t_s}^{(5)}  & -\alpha_{t_s}^{(6)} & -\alpha_{t_s}^{(7)} \\
\alpha_{t_s}^{(1)} & \alpha_{t_s}^{(0)}  & -\alpha_{t_s}^{(3)} & \alpha_{t_s}^{(2)} & -\alpha_{t_s}^{(5)} & \alpha_{t_s}^{(4)}  & \alpha_{t_s}^{(7)} & -\alpha_{t_s}^{(6)} \\
\alpha_{t_s}^{(2)} & \alpha_{t_s}^{(3)}  & \alpha_{t_s}^{(0)} & -\alpha_{t_s}^{(1)} & -\alpha_{t_s}^{(6)} & -\alpha_{t_s}^{(7)}  & \alpha_{t_s}^{(4)} & \alpha_{t_s}^{(5)} \\
\alpha_{t_s}^{(3)} & -\alpha_{t_s}^{(2)}  & \alpha_{t_s}^{(1)} & \alpha_{t_s}^{(0)} & -\alpha_{t_s}^{(7)} & \alpha_{t_s}^{(6)}  & -\alpha_{t_s}^{(5)} & \alpha_{t_s}^{(4)} \\
\alpha_{t_s}^{(4)} & \alpha_{t_s}^{(5)}  & \alpha_{t_s}^{(6)} & \alpha_{t_s}^{(7)} & \alpha_{t_s}^{(0)} & -\alpha_{t_s}^{(1)}  & -\alpha_{t_s}^{(2)} & -\alpha_{t_s}^{(3)} \\
\alpha_{t_s}^{(5)} & -\alpha_{t_s}^{(4)}  & \alpha_{t_s}^{(7)} & -\alpha_{t_s}^{(6)} & \alpha_{t_s}^{(1)} & \alpha_{t_s}^{(0)}  & \alpha_{t_s}^{(3)} & -\alpha_{t_s}^{(2)} \\
\alpha_{t_s}^{(6)} & -\alpha_{t_s}^{(7)}  & -\alpha_{t_s}^{(4)} & \alpha_{t_s}^{(5)} & \alpha_{t_s}^{(2)} & -\alpha_{t_s}^{(3)}  & \alpha_{t_s}^{(0)} & \alpha_{t_s}^{(1)} \\
\alpha_{t_s}^{(7)} & \alpha_{t_s}^{(6)}  & -\alpha_{t_s}^{(5)} & -\alpha_{t_s}^{(4)} & \alpha_{t_s}^{(3)} & \alpha_{t_s}^{(2)}  & -\alpha_{t_s}^{(1)} & \alpha_{t_s}^{(0)} 
 \end{array} \right)\end{equation} 

the model on octonions can be represented as an eight-dimensional model 

\begin{equation}
\bm{y}_{t_j}=F_{t_j}\bm{y}_{t_{j-1}} +\bm{\epsilon_{t_{j}}}\label{eq:sstransitionO}
\end{equation}

And Eq. (\ref{eq:sstransitionO}) can be considered as the transition equation in the state space representation of the model. To complete the state-space representation of the model in the $\mathbb{C}$, $\mathbb{H}$ and $\mathbb{O}$, the observation equation needs to be specified. This equation can be written similarly as Eq. (\ref{eq3}), (\ref{eq4}) or (\ref{eq5}),

\begin{equation}
\bm{y}_{t_j}^{\ast} =G_{t_j}\bm{y}_{t_{j}}+\delta W_{t_j}\label{eqObs}
\end{equation}

\noindent
where $\delta$ is equal to $1$ when there is measurement error in the observations, and zero otherwise. The matrix $G_{t_j}$ can be the identity matrix if all the coordinates of the multivariate model are observed (as in Eqs. (\ref{eq4})) or (\ref{eq5})), or can be a smaller matrix if only some coordinates are observed as in Eq. (\ref{eq3}).

\subsection{ Proof of Corollary 1}
\label{sec:proofc1}
By Theorem $2$, the process described by Equation (\ref{eq:itsMoivre}) can be represented as a state-space system as follows:

\begin{eqnarray}
\bm{x}_{t_j} &=&F_{t_j}\bm{x}_{t_{j-1}}+V_{t_j}\\
\bm{y}_{t_j} &=&G_{t_j}\bm{x}_{t_{j}}+W_{t_j} \mbox{ j=1,\ldots,n}.
\end{eqnarray}

When the random variables are quaternions, the transition matrices $F_{t_s}$ are defined as

\begin{equation}
F_{t_s} = \left(\begin{array}{cccc} \alpha_{t_s}^{(0)} & -\alpha\alpha_{t_s}^{(1)}  & -\beta\alpha_{t_s}^{(2)} & -\alpha\beta\alpha_{t_s}^{(3)} \\ 
\alpha_{t_s}^{(1)} & \alpha_{t_s}^{(0)} & -\beta\alpha_{t_s}^{(3)} & \beta\alpha_{t_s}^{(2)} \\ 
\alpha_{t_s}^{(2)} & \alpha\alpha_{t_s}^{(3)} & \alpha_{t_s}^{(0)} & -\alpha\alpha_{t_s}^{(1)}  \\ 
\alpha_{t_s}^{(3)} & -\alpha_{t_s}^{(2)} & \alpha_{t_s}^{(1)}  & \alpha_{t_s}^{(0)} \end{array} \right),\end{equation}

\noindent 
and the observational matrices $G_{t_s}$ can take different forms depending on the dimension of the multivariate time series. For example, for a three-dimensional time series $G_{t_s}$ can take the following forms:

\begin{equation}
G_{t_s} = \left(\begin{array}{cccc} 1 & 0 & 0 & 0 \\
0 & 1 & 0 & 0 \\
0 & 0 & 1 & 0 \\
\end{array} \right)\end{equation}

\begin{equation}
G_{t_s} = \left(\begin{array}{cccc} 1 & 0 & 0 & 0 \\
0 & 1 & 0 & 0 \\
0 & 0 & 0 & 1 \\
\end{array} \right)\end{equation}

\begin{equation}
G_{t_s} = \left(\begin{array}{cccc} 1 & 0 & 0 & 0 \\
0 & 0 & 1 & 0 \\
0 & 0 & 0 & 1 \\
\end{array} \right)\end{equation}

\begin{equation}
G_{t_s} = \left(\begin{array}{cccc} 0 & 1 & 0 & 0 \\
0 & 0 & 1 & 0 \\
0 & 0 & 0 & 1 \\
\end{array} \right).\end{equation}

For two-dimensional time series $G_{t_s}$ can take the following forms:

\begin{equation}
G_{t_s} = \left(\begin{array}{cccc} 1 & 0 & 0 & 0 \\
0 & 1 & 0 & 0 \\
\end{array} \right)\end{equation}

\begin{equation}
G_{t_s} = \left(\begin{array}{cccc} 1 & 0 & 0 & 0 \\
0 & 0 & 1 & 0 \\
\end{array} \right)\end{equation}

\begin{equation}
G_{t_s} = \left(\begin{array}{cccc} 1 & 0 & 0 & 0 \\
0 & 0 & 0 & 1 \\
\end{array} \right)\end{equation}

\begin{equation}
G_{t_s} = \left(\begin{array}{cccc} 0 & 1 & 0 & 0 \\
0 & 0 & 1 & 0 \\
\end{array} \right)\end{equation}

\begin{equation}
G_{t_s} = \left(\begin{array}{cccc} 0 & 1 & 0 & 0 \\
0 & 0 & 0 & 1 \\
\end{array} \right)\end{equation}

\begin{equation}
G_{t_s} = \left(\begin{array}{cccc} 0 & 0 & 1 & 0 \\
0 & 0 & 0 & 1 \\
\end{array} \right)\end{equation}

The observability matrix corresponds to $ G_{t_j}\times F_{t_j}$. This matrix needs to be full rank for the process to be defined.

\subsection{ Proof of Theorem 3}


Consider the model defined as

\begin{eqnarray}
    \bm{y}_{t_j}&=&\bm{\phi}^{t_j-t_{j-1}}\bm{y}_{t_{j-1}} +\bm{\epsilon_{t_{j}}}\label{eqset1}
\end{eqnarray}

    We will show that $\bm{y}_{t_j}=\sum_{k=0}^{\infty}\bm{\phi}^{t_j-t_{j-k}}\bm{\epsilon}_{t_{j-k}}$ is a solution of the model described by equation (\ref{eqset1}).
    \begin{eqnarray}
    \bm{\phi}^{t_j-t_{j-1}}\bm{y}_{t_{j-1}}+ \bm{\epsilon_{t_{j}}} &=&
   \bm{\phi}^{t_j-t_{j-1}}(\sum_{k=0}^{\infty}\bm{\phi}^{t_{j-1}-t_{j-1-k}}\bm{\epsilon}_{t_{j-1-k}})+\bm{\epsilon_{t_{j}}}\\ \label{eqset2}
    &=& \sum_{k=0}^{\infty}\bm{\phi}^{t_j-t_{j-1}}\bm{\phi}^{t_{j-1}-t_{j-1-k}}\bm{\epsilon}_{t_{j-1-k}}+\bm{\epsilon_{t_{j}}}\\ \label{eqset3}
    &=& \sum_{k=0}^{\infty}\bm{\phi}^{t_j-t_{j-1-k}}\bm{\epsilon}_{t_{j-1-k}}+\bm{\epsilon_{t_{j}}}\\ \label{eqset4}
    &=& \sum_{k=0}^{\infty}\bm{\phi}^{t_j-t_{j-k}}\bm{\epsilon}_{t_{j-k}}\\ \label{eqset5}
    &=&  \bm{y}_{t_j}
\end{eqnarray}

From Eq. (\ref{eqset2}) to (\ref{eqset3}), it has been used the property of  associativity of the multiplication. From Eq. (\ref{eqset3}) to (\ref{eqset4}), it has been used the polar representation of complex and octonions and the De Moivre's formula. 

From Eq. (\ref{eqset1}) we can derive the variance of the process as follows
\begin{eqnarray}
    V(\bm{y}_{t_j})&=& V(\sum_{k=0}^{\infty}\bm{\phi}^{t_j-t_{j-k}}\bm{\epsilon_{t_{j-k}}} )\\
                   &=& \sum_{k=0}^{\infty}\bm{\phi}^{t_j-t_{j-k}}V(\bm{\epsilon_{t_{j-k}}})\bm{\phi}^{\ast (t_j-t_{j-k})} \\
                   &=& \sum_{k=0}^{\infty}\bm{\phi}^{t_j-t_{j-k}}(\Sigma-\phi^{t_{j-k}-t_{j-k-1}}\Sigma\phi^{\ast(t_{j-k}-t_{j-k-1})})\bm{\phi}^{\ast (t_j-t_{j-k})} \\
                   &=& \sum_{k=0}^{\infty}\bm{\phi}^{t_j-t_{j-k}}\Sigma\bm{\phi}^{\ast (t_j-t_{j-k})} -\\
                   &&\bm{\phi}^{t_j-t_{j-k}}\phi^{t_{j-k}-t_{j-k-1}}\Sigma\phi^{\ast(t_{j-k}-t_{j-k-1})}\bm{\phi}^{\ast (t_j-t_{j-k})} \\
                   &=& \sum_{k=0}^{\infty}\bm{\phi}^{t_j-t_{j-k}}\Sigma\bm{\phi}^{\ast (t_j-t_{j-k})} -\bm{\phi}^{t_j-t_{j-k-1}}\Sigma\phi^{\ast(t_{j}-t_{j-k-1})} \\
                   &=& \Sigma -\bm{\phi}^{t_j-t_{j-1}}\Sigma\phi^{\ast(t_{j}-t_{j-1})}+ \\
                   && \bm{\phi}^{t_j-t_{j-1}}\Sigma\bm{\phi}^{\ast (t_j-t_{j-1})} -\bm{\phi}^{t_j-t_{j-2}}\Sigma\phi^{\ast(t_{j}-t_{j-2})} +\\
                   && \bm{\phi}^{t_j-t_{j-2}}\Sigma\bm{\phi}^{\ast (t_j-t_{j-2})} -\bm{\phi}^{t_j-t_{j-3}}\Sigma\phi^{\ast(t_{j}-t_{j-3})} + \ldots=\Sigma
\end{eqnarray}

Here $\phi^{\ast}$ represents the conjugate of $\phi$.



\subsection{Proof of Theorem 4}

A process represented by a linear state-space model is stationary (or more precisely, covariance stationary) if the system is stable, meaning the state transition matrix is ``causal'' or ``stable'' ensuring that the influence of past shocks decays over time. Specifically, for a discrete-time linear state-space model (\(X_{t+1}=FX_{t}+w_{t+1}\)), the process is stationary if all eigenvalues of the state transition matrix \(F\) have absolute values less than one. For autoregressive processes, it is given that each coordinate $\phi_i$ of the parameter vector $\mathbf{\phi}$ has absolute value less than 1. Therefore, this condition together with the condition that $\mbox{det}(F_{t_s})<1$ defines the parameter space where the process is stationary.\\

\noindent
{\bf BiAR}
\begin{equation}
F_{t_s} = \left(\begin{array}{cc} \alpha^{(0)}_{t_s} & -\alpha^{(1)}_{t_s}\\\alpha^{(1)}_{t_s}&\alpha^{(0)}_{t_s}
\end{array}\right)\end{equation}

\noindent
where $\alpha^{(0)}_{t_s}=\|\mathbf{\phi}\|^{\delta_s}\cos(\delta_s\psi)$ and $\alpha^{(1)}_{t_s}=\|\mathbf{\phi}\|^{\delta_s}\sin(\delta_s\psi)$.

The $\mbox{det}(F_{t_s})=(\alpha^{(0)}_{t_s})^2+(\alpha^{(1)}_{t_s})^2=\|\mathbf{\phi}\|^{2\delta_s}=(\phi_0^2+\phi_1^2)^{\delta_s}$.\\

The eigenvalues of the transition matrix $F_{t_s}$ are: $\lambda_1=\|\mathbf{\phi}\|^{\delta_s}\cos(\delta_s\psi)-\sqrt{-\|\mathbf{\phi}\|^{2\delta_s}\sin^2(\delta_s\psi)}$ and $\lambda_2=\|\mathbf{\phi}\|^{\delta_s}\cos(\delta_s\psi)+\sqrt{-\|\mathbf{\phi}\|^{2\delta_s}\sin^2(\delta_s\psi)}$.
\vspace{0.3in}

The process is stationary if the eigenvalues are less than 1 which implies that the $\mbox{det}(F_{t_s})<1$ for all $\delta_s>0$. This is equivalent to assume that $\phi_0^2+\phi_1^2<1$.
\vspace{0.5in}

\noindent
{\bf Generalized BiAR}

\begin{equation}
F_{t_s} = \left(\begin{array}{cc} \alpha^{(0)}_{t_s} & -\alpha\alpha^{(1)}_{t_s}\\\alpha^{(1)}_{t_s}&\alpha^{(0)}_{t_s}
\end{array}\right)\end{equation}

\noindent
where $\alpha^{(0)}_{t_s}=\|\mathbf{\phi}\|_L^{\delta_s}\cos(\delta_s\psi)$ and $\alpha^{(1)}_{t_s}=\|\mathbf{\phi}\|_L^{\delta_s}\sin(\delta_s\psi)$.

The $\mbox{det}(F_{t_s})=(\alpha^{(0)}_{t_s})^2-\alpha(\alpha^{(1)}_{t_s})^2$.\\

The eigenvalues of the transition matrix $F_{t_s}$ are: $\lambda_1=\alpha^{(0)}_{t_s}-\sqrt{\alpha}\alpha^{(1)}_{t_s}$ and $\lambda_2=\alpha^{(0)}_{t_s}+\sqrt{\alpha}\alpha^{(1)}_{t_s}$.
\vspace{0.3in}

The process is stationary if the eigenvalues are less than 1 which implies that the $\mbox{det}(F_{t_s})<1$ for all $\delta_s>0$. 
\vspace{0.5in}

\noindent
{\bf QiAR}
\begin{equation}
F_{t_s} = \left(\begin{array}{cccc} \alpha_{t_s}^{(0)} & -\alpha_{t_s}^{(1)}  & -\alpha_{t_s}^{(2)} & -\alpha_{t_s}^{(3)} \\ \alpha_{t_s}^{(1)} & \alpha_{t_s}^{(0)} & -\alpha_{t_s}^{(3)} & \alpha_{t_s}^{(2)} \\ \alpha_{t_s}^{(2)} & \alpha_{t_s}^{(3)} & \alpha_{t_s}^{(0)} & -\alpha_{t_s}^{(1)}  \\ \alpha_{t_s}^{(3)} & -\alpha_{t_s}^{(2)} & \alpha_{t_s}^{(1)}  & \alpha_{t_s}^{(0)} \end{array} \right)\end{equation} 

\noindent
where $\alpha^{(0)}_{t_s}=\|\mathbf{\phi}\|^{\delta_s}\cos(\delta_s\psi)$, $\alpha^{(1)}_{t_s}=\|\mathbf{\phi}\|^{\delta_s}\frac{\phi_1}{\sqrt{\phi_1^2+\phi_2^2+\phi_3^2}}\sin(\delta_s\psi)$, $\alpha^{(2)}_{t_s}=\|\mathbf{\phi}\|^{\delta_s}\frac{\phi_2}{\sqrt{\phi_1^2+\phi_2^2+\phi_3^2}}\sin(\delta_s\psi)$ and $\alpha^{(3)}_{t_s}=\|\mathbf{\phi}\|^{\delta_s}\frac{\phi_3}{\sqrt{\phi_1^2+\phi_2^2+\phi_3^2}}\sin(\delta_s\psi)$.

The determinant \begin{eqnarray} \mbox{det}(F_{t_s})&=&((\alpha^{(0)}_{t_s})^2+(\alpha^{(1)}_{t_s})^2+(\alpha^{(2)}_{t_s})^2+(\alpha^{(3)}_{t_s})^2)^2\\
&=&\|\mathbf{\phi}\|^{4\delta_s}=(\phi_0^2+\phi_1^2+\phi_2^2+\phi_3^2)^{2\delta_s}.
\end{eqnarray}

The eigenvalues of the transition matrix $F_{t_s}$ are: $\lambda_1=\|\mathbf{\phi}\|^{\delta_s}\cos(\delta_s\psi)-\sqrt{-\|\mathbf{\phi}\|^{2\delta_s}\sin^2(\delta_s\psi)}$ and $\lambda_2=\|\mathbf{\phi}\|^{\delta_s}\cos(\delta_s\psi)+\sqrt{-\|\mathbf{\phi}\|^{2\delta_s}\sin^2(\delta_s\psi)}$; each with multiplicity $2$.
\vspace{0.3in}

The process is stable if the eigenvalues are less than 1 which implies that the $\mbox{det}(F_{t_s})<1$ for all $\delta_s>0$. This is equivalent to assume that $\phi_0^2+\phi_1^2+\phi_2^2+\phi_3^2<1$
\vspace{0.5in}

\noindent
{\bf Generalized QiAR}
\begin{equation}
F_{t_s} = \left(\begin{array}{cccc} \alpha_{t_s}^{(0)} & -\alpha\alpha_{t_s}^{(1)}  & -\beta\alpha_{t_s}^{(2)} & -\alpha\beta\alpha_{t_s}^{(3)} \\ 
\alpha_{t_s}^{(1)} & \alpha_{t_s}^{(0)} & -\beta\alpha_{t_s}^{(3)} & \beta\alpha_{t_s}^{(2)} \\ 
\alpha_{t_s}^{(2)} & \alpha\alpha_{t_s}^{(3)} & \alpha_{t_s}^{(0)} & -\alpha\alpha_{t_s}^{(1)}  \\ 
\alpha_{t_s}^{(3)} & -\alpha_{t_s}^{(2)} & \alpha_{t_s}^{(1)}  & \alpha_{t_s}^{(0)} \end{array} \right),\end{equation}

\noindent
where $\alpha^{(0)}_{t_s}=\|\mathbf{\phi}\|_L^{\delta_s}\cos(\delta_s\psi)$, $\alpha^{(1)}_{t_s}=\|\mathbf{\phi}\|_L^{\delta_s}\frac{\phi_1}{\sqrt{\alpha\phi_1^2+\beta\phi_2^2+\alpha\beta\phi_3^2}}\sin(\delta_s\psi)$, $\alpha^{(2)}_{t_s}=\|\mathbf{\phi}\|_L^{\delta_s}\frac{\phi_2}{\sqrt{\alpha\phi_1^2+\beta\phi_2^2+\alpha\beta\phi_3^2}}\sin(\delta_s\psi)$ and $\alpha^{(3)}_{t_s}=\|\mathbf{\phi}\|_L^{\delta_s}\frac{\phi_3}{\sqrt{\alpha\phi_1^2+\beta\phi_2^2+\alpha\beta\phi_3^2}}\sin(\delta_s\psi)$.

The determinant \begin{eqnarray} \mbox{det}(F_{t_s}) &=&((\alpha^{(0)}_{t_s})^2+\alpha(\alpha^{(1)}_{t_s})^2+\beta(\alpha^{(2)}_{t_s})^2+\alpha\beta(\alpha^{(3)}_{t_s})^2)^2\\
    &=&\|\mathbf{\phi}\|_L^{4\delta_s}=(\phi_0^2+\alpha\phi_1^2+\beta\phi_2^2+\alpha\beta\phi_3^2)^{2\delta_s}.
\end{eqnarray}

The eigenvalues of the transition matrix $F_{t_s}$ are: $\lambda_1=\|\mathbf{\phi}\|_L^{\delta_s}\cos(\delta_s\psi)-\sqrt{-\|\mathbf{\phi}\|_L^{2\delta_s}\sin^2(\delta_s\psi)}$ and $\lambda_2=\|\mathbf{\phi}\|_L^{\delta_s}\cos(\delta_s\psi)+\sqrt{-\|\mathbf{\phi}\|_L^{2\delta_s}\sin^2(\delta_s\psi)}$; each with multiplicity $2$.

The process is stable if the eigenvalues are less than 1 which implies that the $\mbox{det}(F_{t_s})<1$ for all $\delta_s>0$. This is equivalent to assume that $\phi_0^2+\alpha\phi_1^2+\beta\phi_2^2+\alpha\beta\phi_3^2<1$
\vspace{0.5in}

\noindent
{\bf OiAR}
\begin{equation}
F_{t_s} = \left(\begin{array}{cccccccc} 
\alpha_{t_s}^{(0)} & -\alpha_{t_s}^{(1)}  & -\alpha_{t_s}^{(2)} & -\alpha_{t_s}^{(3)} & -\alpha_{t_s}^{(4)} & -\alpha_{t_s}^{(5)}  & -\alpha_{t_s}^{(6)} & -\alpha_{t_s}^{(7)} \\
\alpha_{t_s}^{(1)} & \alpha_{t_s}^{(0)}  & \alpha_{t_s}^{(3)} & -\alpha_{t_s}^{(2)} & \alpha_{t_s}^{(5)} & -\alpha_{t_s}^{(4)}  & -\alpha_{t_s}^{(7)} & \alpha_{t_s}^{(6)} \\
\alpha_{t_s}^{(2)} & -\alpha_{t_s}^{(3)}  & \alpha_{t_s}^{(0)} & \alpha_{t_s}^{(1)} & \alpha_{t_s}^{(6)} & \alpha_{t_s}^{(7)}  & -\alpha_{t_s}^{(4)} & -\alpha_{t_s}^{(5)} \\
\alpha_{t_s}^{(3)} & \alpha_{t_s}^{(2)}  & -\alpha_{t_s}^{(1)} & \alpha_{t_s}^{(0)} & \alpha_{t_s}^{(7)} & -\alpha_{t_s}^{(6)}  & \alpha_{t_s}^{(5)} & -\alpha_{t_s}^{(4)} \\
\alpha_{t_s}^{(4)} & -\alpha_{t_s}^{(5)}  & -\alpha_{t_s}^{(6)} & -\alpha_{t_s}^{(7)} & \alpha_{t_s}^{(0)} & \alpha_{t_s}^{(1)}  & \alpha_{t_s}^{(2)} & \alpha_{t_s}^{(3)} \\
\alpha_{t_s}^{(5)} & \alpha_{t_s}^{(4)}  & -\alpha_{t_s}^{(7)} & \alpha_{t_s}^{(6)} & -\alpha_{t_s}^{(1)} & \alpha_{t_s}^{(0)}  & -\alpha_{t_s}^{(3)} & \alpha_{t_s}^{(2)} \\
\alpha_{t_s}^{(6)} & \alpha_{t_s}^{(7)}  & \alpha_{t_s}^{(4)} & -\alpha_{t_s}^{(5)} & -\alpha_{t_s}^{(2)} & \alpha_{t_s}^{(3)}  & \alpha_{t_s}^{(0)} & -\alpha_{t_s}^{(1)} \\
\alpha_{t_s}^{(7)} & -\alpha_{t_s}^{(6)}  & \alpha_{t_s}^{(5)} & \alpha_{t_s}^{(4)} & -\alpha_{t_s}^{(3)} & -\alpha_{t_s}^{(2)}  & \alpha_{t_s}^{(1)} & \alpha_{t_s}^{(0)} 
 \end{array} \right)\end{equation} 
\vspace{0.3in}

 \noindent
 where $\alpha^{(0)}_{t_s}=\|\mathbf{\phi}\|^{\delta_s}\cos(\delta_s\psi)$, $\alpha^{(i)}_{t_s}=\|\mathbf{\phi}\|^{\delta_s}\frac{\phi_i}{\sqrt{\phi_1^2+\phi_2^2+\phi_3^2+\phi_4^2+\phi_5^2+\phi_6^2+\phi_7^2}}\sin(\delta_s\psi)$ for $\mbox{ i=1,\ldots,7}$.

 The determinant \begin{eqnarray} \mbox{det}(F_{t_s}) &=&((\alpha^{(0)}_{t_s})^2+(\alpha^{(1)}_{t_s})^2+(\alpha^{(2)}_{t_s})^2+(\alpha^{(3)}_{t_s})^2+(\alpha^{(4)}_{t_s})^2+(\alpha^{(5)}_{t_s})^2\\
& & +(\alpha^{(6)}_{t_s})^2+(\alpha^{(7)}_{t_s})^2)^4\\
    &=&\|\mathbf{\phi}\|^{4\delta_s}
\end{eqnarray}

The eigenvalues of the transition matrix $F_{t_s}$ are: 
$$\lambda_1=\alpha^{(0)}-\sqrt{-(\alpha^{(1)}_{t_s})^2-(\alpha^{(2)}_{t_s})^2-(\alpha^{(3)}_{t_s})^2-(\alpha^{(4)}_{t_s})^2-(\alpha^{(5)}_{t_s})^2 -(\alpha^{(6)}_{t_s})^2-(\alpha^{(7)}_{t_s})^2}$$ $$\lambda_2=\alpha^{(0)}+\sqrt{-(\alpha^{(1)}_{t_s})^2-(\alpha^{(2)}_{t_s})^2-(\alpha^{(3)}_{t_s})^2-(\alpha^{(4)}_{t_s})^2-(\alpha^{(5)}_{t_s})^2 -(\alpha^{(6)}_{t_s})^2-(\alpha^{(7)}_{t_s})^2},$$ each with multiplicity $4$. These are equivalent to:
$\lambda_1=\|\mathbf{\phi}\|^{\delta_s}\cos(\delta_s\psi)-\sqrt{-\|\mathbf{\phi}\|^{2\delta_s}\sin^2(\delta_s\psi)}$ and $\lambda_2=\|\mathbf{\phi}\|^{\delta_s}\cos(\delta_s\psi)+\sqrt{-\|\mathbf{\phi}\|^{2\delta_s}\sin^2(\delta_s\psi)}$; each with multiplicity $4$.
\vspace{0.5in}

The process is stable if the eigenvalues are less than 1 which implies that the $\mbox{det}(F_{t_s})<1$ for all $\delta_s>0$. This is equivalent to assume that $\|\mathbf{\phi}\|^{4}<1$
\vspace{0.5in}

 \noindent
{\bf Generalized OiAR}

\begin{equation}
F_{t_s} = \left(\begin{array}{cccccccc} 
\alpha_{t_s}^{(0)} & -\alpha\alpha_{t_s}^{(1)}  & -\beta\alpha_{t_s}^{(2)} & -\alpha\beta\alpha_{t_s}^{(3)} & -\gamma\alpha_{t_s}^{(4)} & -\alpha\gamma\alpha_{t_s}^{(5)}  & -\beta\gamma\alpha_{t_s}^{(6)} & -\alpha\beta\gamma\alpha_{t_s}^{(7)} \\
\alpha_{t_s}^{(1)} & \alpha_{t_s}^{(0)}  & \beta\alpha_{t_s}^{(3)} & -\beta\alpha_{t_s}^{(2)} & \gamma\alpha_{t_s}^{(5)} & -\gamma\alpha_{t_s}^{(4)}  & -\beta\gamma\alpha_{t_s}^{(7)} & \beta\gamma\alpha_{t_s}^{(6)} \\
\alpha_{t_s}^{(2)} & -\alpha\alpha_{t_s}^{(3)}  & \alpha_{t_s}^{(0)} & \alpha\alpha_{t_s}^{(1)} & \gamma\alpha_{t_s}^{(6)} & \alpha\gamma\alpha_{t_s}^{(7)}  & -\gamma\alpha_{t_s}^{(4)} & -\alpha\gamma\alpha_{t_s}^{(5)} \\
\alpha_{t_s}^{(3)} & \alpha_{t_s}^{(2)}  & -\alpha_{t_s}^{(1)} & \alpha_{t_s}^{(0)} & \gamma\alpha_{t_s}^{(7)} & -\gamma\alpha_{t_s}^{(6)}  & \gamma\alpha_{t_s}^{(5)} & -\gamma\alpha_{t_s}^{(4)} \\
\alpha_{t_s}^{(4)} & -\alpha\alpha_{t_s}^{(5)}  & -\beta\alpha_{t_s}^{(6)} & -\alpha\beta\alpha_{t_s}^{(7)} & \alpha_{t_s}^{(0)} & \alpha\alpha_{t_s}^{(1)}  & \beta\alpha_{t_s}^{(2)} & \alpha\beta\alpha_{t_s}^{(3)} \\
\alpha_{t_s}^{(5)} & \alpha_{t_s}^{(4)}  & -\beta\alpha_{t_s}^{(7)} & \beta\alpha_{t_s}^{(6)} & -\alpha_{t_s}^{(1)} & \alpha_{t_s}^{(0)}  & -\beta\alpha_{t_s}^{(3)} & \beta\alpha_{t_s}^{(2)} \\
\alpha_{t_s}^{(6)} & \alpha\alpha_{t_s}^{(7)}  & \alpha_{t_s}^{(4)} & -\alpha\alpha_{t_s}^{(5)} & -\alpha_{t_s}^{(2)} & \alpha\alpha_{t_s}^{(3)}  & \alpha_{t_s}^{(0)} & -\alpha\alpha_{t_s}^{(1)} \\
\alpha_{t_s}^{(7)} & -\alpha_{t_s}^{(6)}  & \alpha_{t_s}^{(5)} & \alpha_{t_s}^{(4)} & -\alpha_{t_s}^{(3)} & -\alpha_{t_s}^{(2)}  & \alpha_{t_s}^{(1)} & \alpha_{t_s}^{(0)} 
 \end{array} \right)\end{equation} 
\vspace{0.3in}

The determinant \begin{eqnarray} \mbox{det}(F_{t_s}) &=&((\alpha^{(0)}_{t_s})^2+\alpha(\alpha^{(1)}_{t_s})^2+\beta(\alpha^{(2)}_{t_s})^2+\alpha\beta(\alpha^{(3)}_{t_s})^2+\gamma(\alpha^{(4)}_{t_s})^2\\
& &+\alpha\gamma(\alpha^{(5)}_{t_s})^2+\beta\gamma(\alpha^{(6)}_{t_s})^2+\alpha\beta\gamma(\alpha^{(7)}_{t_s})^2)^4\\
    &=&\|\mathbf{\phi}\|_L^{4\delta_s}
\end{eqnarray}

The eigenvalues of the transition matrix $F_{t_s}$ are:
$$\lambda_1=(\alpha^{(0)}_{t_s})+\sqrt{-\alpha(\alpha^{(1)}_{t_s})^2-\beta(\alpha^{(2)}_{t_s})^2-\alpha\beta(\alpha^{(3)}_{t_s})^2-\gamma(\alpha^{(4)}_{t_s})^2-\alpha\gamma(\alpha^{(5)}_{t_s})^2-\beta\gamma(\alpha^{(6)}_{t_s})^2-\alpha\beta\gamma(\alpha^{(7)}_{t_s})^2}$$ $$\lambda_2=(\alpha^{(0)}_{t_s})-\\\sqrt{-\alpha(\alpha^{(1)}_{t_s})^2-\beta(\alpha^{(2)}_{t_s})^2-\alpha\beta(\alpha^{(3)}_{t_s})^2-\gamma(\alpha^{(4)}_{t_s})^2-\alpha\gamma(\alpha^{(5)}_{t_s})^2-\beta\gamma(\alpha^{(6)}_{t_s})^2-\alpha\beta\gamma(\alpha^{(7)}_{t_s})^2}$$
$\lambda_1=\|\mathbf{\phi}\|_L^{\delta_s}\cos(\delta_s\psi)-\sqrt{-\|\mathbf{\phi}\|_L^{2\delta_s}\sin^2(\delta_s\psi)}$ and $\lambda_2=\|\mathbf{\phi}\|_L^{\delta_s}\cos(\delta_s\psi)+\sqrt{-\|\mathbf{\phi}\|_L^{2\delta_s}\sin^2(\delta_s\psi)}$; each with multiplicity $4$.
\vspace{0.5in}

The process is stable if  if the eigenvalues are less than 1 which implies that the $\mbox{det}(F_{t_s})<1$ for all $\delta_s>0$. This is equivalent to assume that $\|\mathbf{\phi}\|_L^{4}<1$
\vspace{0.5in}

\subsection{Basic definition and properties of the Complex ($\mathbb{C}$), Quaternions ($\mathbb{H}$) and Octonions ($\mathbb{O}$) algebras}

Quaternions and octonions are two of the four normed division algebras that extend the familiar concepts of real and complex numbers\cite{conway2003}. With addition and quaternionic/octonionic multiplication, these form vector spaces are algebras. Having a conjugation map, it is possible also to define a norm. The quaternions are non-commutative, and the octonions are non-commutative and non-associative, but they satisfy the properties of being alternate (e.g $x^2y=x(xy)$ and $yx^2=(yx)x$), flexible (e.g $x(yx)=(xy)x$) and power-associative (e.g for each $x$, the subalgebra generated by $x$ is an associative algebra) \cite{ahmed2021powers,jafari2015generalized}.

\subsubsection{The Complex ($\mathbb{C}$)}
$x\in \mathbb{C}$ if $x$ can be written as $x=a_0+ia_1$ where $a_l\in \mathbb{R}\mbox{ with l=0,1}$ and $i^2=-1$. $\mathbb{R}$ represents the real numbers.\\

\noindent
{\bf Polar representation:} $x\in \mathbb{C}$ with $x=a_0+ia_1$, then $x$ has a polar representation such that $x=\|x\|(\cos(\theta)+i\sin(\theta))$ where $\theta=\tan^{-1}(\lvert a_1\rvert/a_0)$ and $\|x\|=\sqrt{a_0^2+a_1^2}$.\\

\noindent
{\bf Euler's formula:} $\bm{x}$ can be represented as $\bm{x}=\|\bm{x}\|\exp(i\theta)$\\

\noindent
{\bf de Moivre's formula:} $x\in \mathbb{C}$ and $p$ a rational number, then $x^p=\|x\|^p(\cos(p\theta)+i\sin(p\theta))$.

\subsubsection{The generalized Complex ($\mathbb{C_{\alpha}}$)}

$x\in \mathbb{C}$ if $x$ can be written as $x=a_0+ia_1$ where $a_l\in \mathbb{R}\mbox{ with l=0,1}$ and $i^2=\alpha$. $\mathbb{R}$ represents the real numbers.

\noindent
{\bf Polar representation:} $x\in \mathbb{C}$ with $x=a_0+ia_1$, then $x$ has a polar representation such that $x=\|x\|(\cos(\theta)+i\sin(\theta))$ where $\theta=\tan^{-1}(\sqrt{\lvert \alpha a_1^2\rvert}/a_0)$  and $\|x\|=\sqrt{\lvert a_0^2-\alpha a_1^2\rvert}$.\\

\noindent
{\bf Euler's formula:} $\bm{x}$ can be represented as $\bm{x}=\|\bm{x}\|\exp(i\theta)$\\

\noindent
{\bf de Moivre's formula:} $x\in \mathbb{C}$ and $p$ a rational number, then $x^p=\|x\|^p(\cos(p\theta)+i\sin(p\theta))$.

\subsubsection{The Quaternions ($\mathbb{H}$)}
$x\in \mathbb{H}$ if $x$ can be written as $x=a_0+ia_1+ja_2+ka_3$ where $a_l\in \mathbb{R}\mbox{ with l=0,1,2,3}$ and $i^2=-1; j^2=-1; k^2=-1; ij=k; ji=-k$. The multiplication rules is specified in the following table

\begin{table}[ht]
    \centering
    \caption{Quaternion Multiplication Table}
    \begin{tabular}{c|cccc}
        $\cdot$ & $1$ & $i$ & $j$ & $k$ \\
        \hline
        $1$ & $1$ & $i$ & $j$ & $k$ \\
        $i$ & $i$ & $-1$ & $k$ & $-j$ \\
        $j$ & $j$ & $-k$ & $-1$ & $i$ \\
        $k$ & $k$ & $j$ & $-i$ & $-1$ \\
    \end{tabular}
\end{table}

\noindent
{\bf Polar representation:} $x\in \mathbb{H}$ with $x=a_0+ia_1+ja_2+ka_3$, then $x$ has a polar representation such that $x=\|x\|x^{\ast}$ and $x^{\ast}=(\cos(\theta)+w\sin(\theta))$ where $w=\frac{a_1i+a_2j+a_3k}{\sqrt{a_1^2+a_2^2+a_3^2}}$, $\theta=\tan^{-1}(\frac{\sqrt{\lvert a_1^2+a_2^2+a_3^2\rvert}}{a_0})$ and $\|x\|=\sqrt{a_0^2+a_1^2+a_2^2+a_3^2}$.\\

\noindent
{\bf Euler's formula:} $\bm{x}$ can be represented as $\bm{x}=\|\bm{x}\|\exp(w\theta)$\\

\noindent
{\bf de Moivre's formula:} $x\in \mathbb{H}$ and $p$ a rational number, then $x^p=\|x\|^p(\cos(p\theta)+w\sin(p\theta))$.

\subsubsection{The Generalized Quaternions ($\mathbb{H}_{\alpha\beta}$)}
$x\in \mathbb{H}_{\alpha\beta}$ if $x$ can be written as $x=a_0+ia_1+ja_2+ka_3$ where $a_l\in \mathbb{R}\mbox{ with l=0,1,2,3}$ and $i^2=\alpha; j^2=\beta; k^2=-\alpha\beta; ij=k; ji=-k$.The multiplication rules is specified in the following table

\begin{table}[ht]
    \centering
    \caption{Quaternion Multiplication Table}
    \begin{tabular}{c|cccc}
        $\cdot$ & $1$ & $i$ & $j$ & $k$ \\
        \hline
        $1$ & $1$ & $i$ & $j$ & $k$ \\
        $i$ & $i$ & $-\alpha$ & $k$ & $-\alpha j$ \\
        $j$ & $j$ & $-k$ & $-\beta$ & $\beta i$ \\
        $k$ & $k$ & $\alpha j$ & $-\beta i$ & $-\alpha\beta$ \\
    \end{tabular}
\end{table}

\noindent
{\bf Polar representation:} $x\in \mathbb{H}$ with $x=a_0+ia_1+ja_2+ka_3$, then $x$ has a polar representation such that $x=\|x\|(\cos(\theta)+w\sin(\theta))$ where  $\theta=\tan^{-1}(\frac{\sqrt{\lvert \alpha a_1^2+\beta a_2^2+\alpha\beta a_3^2\rvert}}{ a_0})$, $w=\frac{a_1i+a_2j+a_3k}{\sqrt{\alpha a_1^2+\beta a_2^2+\alpha\beta a_3^2}}$, and $\|x\|=\sqrt{\lvert a_0^2+\alpha a_1^2+\beta a_2^2+\alpha\beta a_3^2\rvert}$.\\

\noindent
{\bf Euler's formula:} $\bm{x}$ can be represented as $\bm{x}=\|\bm{x}\|\exp(w\theta)$\\

\noindent
{\bf de Moivre's formula:} $x\in \mathbb{H}$ and $p$ a rational number, then $x^p=\|x\|^p(\cos(p\theta)+w\sin(p\theta))$.

\vspace{0.5cm}
\noindent
\subsubsection{The Octonions ($\mathbb{O}$)}
The octonions can be thought of 8-tuples, so that $x\in \mathbb{O}$ if it can be written as a linear combination of $\{f_0,f_1,f_2,f_3,f_4,f_5,f_6,f_7\}$, i.e. $x=a_0f_0+a_1f_1+a_2f_2+a_3f_3+a_4f_4+a_5f_5+a_6f_6+a_7f_7$ where $a_l\in \mathbb{R}\mbox{ with l=0,1,\ldots,7}$, $f_0=1$ is the scalar and $f_1,f_2,f_3,f_4,f_5,f_6,f_7$ are unit octonions that satisfy the following multiplication rule:

\begin{table}[ht]
    \centering
    \begin{tabular}{c|rrrrrrrr}
        $\cdot$ & $f_0$ & $f_1$ & $f_2$ & $f_3$ & $f_4$ & $f_5$ & $f_6$ & $f_7$ \\
        \hline
        $f_0$ & $f_0$ & $f_1$ & $f_2$ & $f_3$ & $f_4$ & $f_5$ & $f_6$ & $f_7$ \\
        $f_1$ & $f_1$ & $-f_0$ & $f_3$ & $-f_2$ & $f_5$ & $-f_4$ & $-f_7$ & $f_6$ \\
        $f_2$ & $f_2$ & $-f_3$ & $-f_0$ & $f_1$ & $f_6$ & $f_7$ & $-f_4$ & $-f_5$ \\
        $f_3$ & $f_3$ & $f_2$ & $-f_1$ & $-f_0$ & $f_7$ & $-f_6$ & $f_5$ & $-f_4$ \\
        $f_4$ & $f_4$ & $-f_5$ & $-f_6$ & $-f_7$ & $-f_0$ & $f_1$ & $f_2$ & $f_3$ \\
        $f_5$ & $f_5$ & $f_4$ & $-f_7$ & $f_6$ & $-f_1$ & $-f_0$ & $-f_3$ & $f_2$ \\
        $f_6$ & $f_6$ & $f_7$ & $f_4$ & $-f_5$ & $-f_2$ & $f_3$ & $-f_0$ & $-f_1$ \\
        $f_7$ & $f_7$ & $-f_6$ & $f_5$ & $f_4$ & $-f_3$ & $-f_2$ & $f_1$ & $-f_0$ \\
    \end{tabular}
    \caption{Octonion Multiplication Table}
\end{table}

\noindent
{\bf Polar representation:} $x\in \mathbb{O}$ with $x=a_0f_0+a_1f_1+a_2f_2+a_3f_3+a_4f_4+a_5f_5+a_6f_6+a_7f_7$, then $x$ has a polar representation such that $x=\|x\|x^{\ast}$ and $x^{\ast}=(\cos(\theta)+v\sin(\theta))$ where $v=\frac{a_1f_1+a_2f_2+a_3f_3+a_4f_4+a_5f_5+a_6f_6+a_7f_7}{\sqrt{a_1^2+a_2^2+a_3^2+a_4^2+a_5^2+a_6^2+a_7^2}}$, $\theta=\tan^{-1}{(\frac{\sqrt{\lvert a_1^2+a_2^2+a_3^2+a_4^2+a_5^2+a_6^2+a_7^2\rvert}}{a_{0}})}$ and $\|x\|=\sqrt{a_0^2+a_1^2+a_2^2+a_3^2+a_4^2+a_5^2+a_6^2+a_7^2}$.\\

\noindent
{\bf Euler's formula:} $\bm{x}$ can be represented as $\bm{x}=\|\bm{x}\|\exp(i\theta)$\\

\noindent
{\bf de Moivre's formula:} $x\in \mathbb{O}$ and $p$ a rational number, then $x^p=\|x\|^p(\cos(p\theta)+v\sin(p\theta))$.

\subsubsection{The generalized  Octonions ($\mathbb{O}_{\alpha\beta\gamma}$)}

The multiplication rule in this case is specified in the following table:

\begin{table}[ht]
    \centering
    \begin{tabular}{c|rrrrrrrr}
        $\cdot$ & $f_0$ & $f_1$ & $f_2$ & $f_3$ & $f_4$ & $f_5$ & $f_6$ & $f_7$ \\
        \hline
        $f_0$ & $f_0$ & $f_1$ & $f_2$ & $f_3$ & $f_4$ & $f_5$ & $f_6$ & $f_7$ \\
        $f_1$ & $f_1$ & $-\alpha f_0$ & $f_3$ & $-\alpha f_2$ & $f_5$ & $-\alpha f_4$ & $-f_7$ & $\alpha f_6$ \\
        $f_2$ & $f_2$ & $-f_3$ & $-\beta f_0$ & $\beta f_1$ & $f_6$ & $f_7$ & $-\beta f_4$ & $-\beta f_5$ \\
        $f_3$ & $f_3$ & $\alpha f_2$ & $-\beta f_1$ & $-\alpha \beta f_0$ & $f_7$ & $-\alpha f_6$ & $\beta f_5$ & $-\alpha \beta f_4$ \\
        $f_4$ & $f_4$ & $-f_5$ & $-f_6$ & $-f_7$ & $-\gamma f_0$ & $\gamma f_1$ & $\gamma f_2$ & $\gamma f_3$ \\
        $f_5$ & $f_5$ & $\alpha f_4$ & $-f_7$ & $\alpha f_6$ & $-\gamma f_1$ & $-\alpha \gamma f_0$ & $-\gamma f_3$ & $\alpha \gamma f_2$ \\
        $f_6$ & $f_6$ & $f_7$ & $\beta f_4$ & $-\beta f_5$ & $-\gamma f_2$ & $\gamma f_3$ & $-\beta \gamma f_0$ & $-\beta \gamma f_1$ \\
        $f_7$ & $f_7$ & $-\alpha f_6$ & $\beta f_5$ & $\alpha \beta f_4$ & $-\gamma f_3$ & $-\alpha \gamma f_2$ & $\beta \gamma f_1$ & $-\alpha \beta \gamma f_0$ \\
    \end{tabular}
    \caption{Octonion Multiplication Table}
\end{table}

\noindent
{\bf Polar representation:} $x\in \mathbb{O}$ with $x=a_0f_0+a_1f_1+a_2f_2+a_3f_3+a_4f_4+a_5f_5+a_6f_6+a_7f_7$, then $x$ has a polar representation such that $x=\|x\|_Lx^{\ast}$ and $x^{\ast}=(\cos(\theta)+v\sin(\theta))$ where $v=\frac{a_1f_1+a_2f_2+a_3f_3+a_4f_4+a_5f_5+a_6f_6+a_7f_7}{\sqrt{\alpha a_1^2+\beta a_2^2+\alpha\beta a_3^2+\gamma a_4^2+\alpha\gamma a_5^2+\beta\gamma a_6^2+\alpha\beta\gamma a_7^2}}$, $\theta=\tan^{-1}{(\frac{\sqrt{\lvert \alpha a_1^2+\beta a_2^2+\alpha\beta a_3^2+\gamma a_4^2+\alpha\gamma a_5^2+\beta\gamma a_6^2+\alpha\beta\gamma a_7^2\rvert}}{a_{0}})}$ and \\$\|x\|_L=\sqrt{a_0^2+\alpha a_1^2+\beta a_2^2+\alpha\beta a_3^2+\gamma a_4^2+\alpha\gamma a_5^2+\beta\gamma a_6^2+\alpha\beta\gamma a_7^2}$.\\

\noindent
{\bf Euler's formula:} $\bm{x}$ can be represented as $\bm{x}=\|\bm{x}\|_L\exp(i\theta)$\\

\noindent
{\bf de Moivre's formula:} $x\in \mathbb{O}$ and $p$ a rational number, then $x^p=\|x\|_L^p(\cos(p\theta)+v\sin(p\theta))$.

\bibliographystyle{plainnat}
\bibliography{references}

\end{document}